# Spin wave propagation in uniform waveguide: effects, modulation and its application

Lei Zheng, Lichuan Jin, Tianlong Wen, Yulong Liao, Xiaoli Tang, Huaiwu Zhang, Zhiyong Zhong[1]

State Key Laboratory of Electronic Thin Films and Integrated Devices, University of Electronic Science and Technology of China, 611731, Chengdu, China

## Abstract

With the advent of the post-Moore era, researches on beyond-Complementary Metal Oxide Semiconductor (CMOS) approaches have been attracted more and more attention. Magnonics, or spin wave is one of the most promising technology beyond CMOS, which magnons-quanta for spin waves-process the information analogous to electronic charges in electronics. Information transmission by spin waves, which uses the frequency, amplitude and (or) phase to encode information, has a great many of advantages such as extremely low energy loss and wide-band frequency. Moreover, using the nonlinear characteristics of spin waves for information transmission can increase the extra degree of freedom of information. This review provides a tutorial overview over the effects of spin wave propagation and recent research progress in uniform spin wave waveguide. The propagation characteristics of spin waves in uniform waveguides and some special propagation phenomena such as spin wave beam splitting and self-focusing are described by combining experimental phenomena and theoretical formulas. Furthermore, we summarize methods for modulating propagation of spin wave in uniform waveguide, and comment on the advantages and limitations of these methods. The review may promote the development of information transmission technology based on spin waves.

## 1. Introduction

The integrated circuits based on CMOS have achieved great success due to their high speed, low power consumption, and high scalability[1-2]. In the past two decades, FinEFT and Gate-All-Around (GAA) technologies have promoted the process of CMOS based chip process to a new level. In recent years, new processes based on CMOS led by giant companies such as TSMC has entered mass production. However, the integrated circuits based on CMOS technology are constrained by physical limits and power dissipation. The further reduction of channel length will bring more power consumption and introduce some quantum effects[3]. In the future, CMOS scaling is expected to slow down greatly due to cost, power consumption and reliability. In response to the gradual failure of Moore's Law caused by the above reasons, researchers have proposed some solutions such as more Moore, more than Moore and beyond CMOS[3-8]. Among them, more Moore refers to the use of new manufacturing techniques to promote Moore's Law[5]; more than Moore refers to the optimization of integrated circuits by relying on new circuit designs and system algorithms[6]. Beyond CMOS refers to the use of new architectures to achieve information processing, that is, utilizing new information carriers such as spin waves and optoelectronics to achieve information transmission[7-8]. Due to the physical limits and power consumption of CMOS, more Moore and more than Moore can only cope with the short-term development of CMOS, and beyond CMOS is the only way for the development of integrated circuits in the future.

---

[1] Corresponding author: zzy@uestc.edu.cn

Among alternative devices beyond CMOS, spin wave devices, which encode information into the amplitude and phase of the spin wave to propagate, have been identified as the most promising candidate due to their low power consumption and high speed[9]. Spin waves (magnonics) represent a phase-coherent collective oscillation of precessing magnetization vectors in a magnetic medium, which were first proposed by Bloch in 1930[10]. From then on, spin waves are attracting more and more attention from researchers due to their novel characteristics. The most prominent advantage of utilizing spin waves as information encoding is that the energy dissipation is extremely low due to the propagation of spin waves without Joule heating[11-12]. Spin waves have a wide application band which covers the GHz band and even reaches THz band. The minimum wavelength of spin waves can be as low as 10 nm, which is in line with the trend of miniaturization and integration of devices. In addition, information can be encoded into the amplitude and phase of spin waves, which makes the spin waves compute not only by Boolean logic but also by non-Boolean logic such as majority gate. Based on the above novel features, spin wave devices have become a promising competitor for beyond CMOS devices[12].

A typical spin wave device consists of three parts: spin wave source (transmitter), spin wave waveguide and spin wave detector. As the key part of spin wave devices, spin wave waveguide plays the role of propagating and modulating spin wave. The propagation characteristics and phenomena of spin wave in different types of waveguides are different. Typical spin waveguides include uniform waveguides and magnonic crystals. Spin waves propagating in uniform waveguides and magnonic crystals show different propagation characteristics and phenomena. The special propagation characteristics and phenomena that exist during the propagation of spin waves also provide great convenience for the design of spin wave devices. Uniform waveguides generally refer to waveguide with uniform magnetization. The propagation of spin waves in uniform waveguides presents wave characteristics (e.g., dispersion, diffraction and interference) and geometric confinement effects (e.g., localization and quantization)[13-21]. Different from uniform waveguides, magnonic crystals are artificially engineered crystals based on metamaterials. In magnonic crystals, spin waves propagation not only shows the wave characteristics, but also exhibits the characteristics of spin wave forbidden bands[22]. With these characteristics, functions such as spin wave beam splitting, self-focusing, frequency conversion, and wavelength conversion can be realized. These special propagation characteristics and phenomena make the design of spin wave devices more diverse and convenient. Thus, the study on propagation characteristics of spin wave in waveguide is the key to the design of spin wave devices. In last two decades, numerous spin wave devices have been implemented by simulation and experiment (e.g., phase shifters, logic gates, transistors, multiplexers) [23-39]. However, most spin wave devices still have the problems of power consumption and efficiency. How to solve these problems is the key to the development of spin wave devices.

The main goal of this review is to provide an introduction of spin wave theory and challenges for spin wave device design based on uniform waveguide. This review starts with an introduction about spin wave propagation characteristics in uniform waveguide (Sec. II). Subsequently, we combine the contents of Sec. II to introduce modulation methods of spin wave (Sec. III). Finally, Sec. IV concludes the review with an overview of the research on spin wave modulation technology and points out the challenges in implementing and designing high-efficiency spin wave devices.

# 2. Effects of Spin wave propagation in uniform waveguides

This section provides an introduction to spin wave propagation characteristics. We first start by explaining the dispersion relation of spin waves, followed by introduction some important phenomena about spin wave propagation.

## 2.1 Dispersion relations in uniform waveguide

The dispersion relation of spin wave in uniform waveguide is complex. It is mainly related to the following factors: direction and size of wavevector, the relative magnitude of exchange interaction and dipole interaction, shape of waveguides, amplitude of external magnetic field and direction of external magnetic field relative to propagation direction of spin wave[13]. Moreover, magnetocrystalline anisotropy and magnetostriction also can affect the dispersion relationship.

For a uniform waveguide of width *w* and thickness *d*, the total wavevector is shown below

$$k_{total} = \sqrt{k_\perp^2 + k_\parallel^2} \qquad (1)$$

where $k_\perp$ and $k_\parallel$ are the components of the wavevector along the length and width of the waveguide, respectively. Due to the size limitation of waveguide width, $k_\perp$ is expressed as

$$k_\perp = \frac{n\pi}{w} (n = 1,2,3...) \qquad (2)$$

The azimuth of wavevector is as following

$$\theta_k = atan(\frac{k_\perp}{k_\parallel}) = atan(\frac{n\pi}{kw}) \qquad (3)$$

By combining the above equations and simplifying the wavevector $k = k_\parallel$, the dispersion relation can be expressed as follows:

$$\omega(k) = \sqrt{(\omega_0 + \omega_m \lambda_{ex}[k^2 + (\frac{n\pi}{\omega})^2])(\omega_0 + \omega_m \lambda_{ex}[k^2 + (\frac{n\pi}{\omega})^2] + \omega_m F)} \qquad (4)$$

$$F = 1 - g\cos^2(\theta_k - \theta_M) + \frac{\omega_0 g(1-g)\sin^2(\theta_k - \theta_M)}{\omega_0 + \omega_m \lambda_{ex}[k^2 + (\frac{n\pi}{w})^2]} \qquad (5)$$

$$g = 1 - \frac{[1 - exp(-d\sqrt{k^2 + (\frac{n\pi}{w})^2})]}{[d\sqrt{k^2 + (\frac{n\pi}{w})^2}]} \qquad (6)$$

where $\omega_0 = \gamma_0 H_{eff}$ ($H_{eff}$ is the effective field in waveguide), $\omega_m = \gamma_0 M_s$, $\lambda_{ex} = 2\frac{A_{ex}}{(\mu_0 M_s^2)}$ ($A_{ex}$ is the exchange constant of waveguide) and $\theta_M$ is the angle between the magnetization direction (or direction of applied field) and the long axis of the waveguide. It should be noted that in theory, Eq. 4 is tenable only when the waveguide width is wide and $kd < 1$, that is, for pure exchange interaction spin waves. The dipole-exchange interaction will lead to a large deviation, when $kd \geq 1$. In addition, Eq. 4 does not consider the effects of spin pinning on the waveguide surface, high-order thickness modes, and magnetocrystalline anisotropy.

  The demagnetization factor and the spin pinning condition at the edge of the waveguide need to be considered, when the width of the spin wave waveguide reaches the micron level or less. Due to the large length of the waveguide, the demagnetization factor in the length direction can be ignored when the waveguide is magnetized along the long axis, $\theta_M = 0$, i.e. $k \parallel M$. In this configuration, the effective field $H_{eff}$ in the waveguide is equal to the applied field $H_0$. The spin pinning condition is mainly determined by the dipole field, and the spin at the edge of waveguide is usually in a partially pinned state. The pinning state can be described with the effective width $w_{eff}$. There are different ways to define the effective width. The most common definition is the distance between two points corresponding to 10% of the maximum effective field at the center of the waveguide. When magnetizing along the waveguide axis, $w_{eff} \geq w$, there are the following situations: $w_{eff} = w$ represents complete pinning, $w_{eff} > w$ represents partial pinning, and $w_{eff} \to \infty$ represents no pinning at all. The effective width of the waveguide with $\theta_M = 0$ is shown as:

$$w_{eff} = \frac{D_{dip}}{D_{dip} - 2} w \tag{7}$$

where $D_{dip} = 2\pi(w/d)/[1 + 2ln(w/d)]$.

$\theta_M = \frac{\pi}{2}$ means that the magnetization direction of the waveguide is along the tangential direction. The applied magnetic field competes with the shape anisotropy equivalent field. At this time, the magnetization tends to align along the long axis of the waveguide, thereby minimizing the static stray field. Obviously, the effective field (internal magnetic field) acting on the waveguide at this time is smaller than the external field and presents a non-uniform distribution, that is, the effective field at the edge of the waveguide is the smallest, and the effective field at the center of the waveguide is the largest. Assuming that the direction of magnetization in the waveguide is always along the width of the waveguide, the effective field is expressed as:

$$\mu_0 H_{eff} = \mu_0 H_0 - \frac{\mu_0 M_s}{\pi} [a\,tan(\frac{d}{2z + w}) - a\,tan(\frac{d}{2z - w})] \tag{8}$$

where $z$ is the coordinate along the tangential direction of the waveguide and $-w/2 \leq z \leq w/2$. Usually, spin waves can be propagated either along the central area (Central mode) of the waveguide or along the edges (Edge mode) of the waveguide. For the central mode, the effective field of the waveguide is considered to be uniform, and its maximum value is given by Eq. 8 with $z = 0$. Obviously, for the case where the magnetization direction is along the tangential direction of the waveguide, the effective width of the waveguide satisfies $w_{eff} < w$.

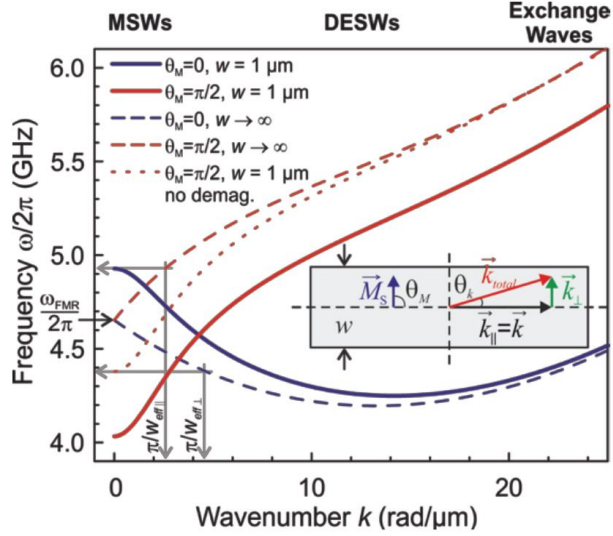

Fig. 1 The dispersion relationship of YIG uniform waveguide and YIG film.

Fig. 1 shows the dispersion relationship of a uniform YIG waveguide and YIG film. The width and thickness are $w = 1\ um$ (waveguide), $w = \infty\ um$ (film), $d = 100\ nm$, respectively. The saturation magnetization and exchange constant are $M_s = 140\ kA/m$, $A_{ex} = 3.5\ pJ/m$, respectively. The applied magnetic field $\mu_0 H_0 = 100\ mT$ is parallel to the waveguide or film plane. Whether in magnetic films or waveguides, the dispersion relation of spin waves can be divided into three regions: the magnetostatic waves with small wavenumber, the exchange waves with large wavenumber and the dipolar-exchange spin wave (DESW) that wavenumber is between the first two. For the spin waves of $\theta_M = 0$ and $\theta_M = \frac{\pi}{2}$ in magnetic films, their frequencies are ferromagnetic resonance frequencies (The corresponding frequency in Fig. 1 is 4.65 GHz) of magnetic films with $k = 0$. However, it's completely different for uniform waveguides. $k_\perp > 0$ always holds in uniform waveguides. For the spin waves of $\theta_M = 0$ (n = 1), calculating from Eq. 7 shows that $w_{eff} \approx 1.22 um$, then $k_\perp = \frac{\pi}{w_{eff}} = 2.58\ rad/um$. Thus, its frequency $f = 4.93 GHz$ is higher than the ferromagnetic resonance frequency when $k = 0$. For the spin waves of $\theta_M = \frac{\pi}{2}$, its frequency $f = 4.03\ GHz$ is lower than the ferromagnetic resonance frequency when $k = 0$. The reason for this phenomenon is that the effective width ($w_{eff} = 0.68 um$) of waveguides becomes smaller and the effective magnetic field on waveguides is smaller than the external magnetic field because of the existence of demagnetization field.

## 2.2 Special phenomena of spin wave propagation in uniform waveguides

### 2.2.1 Channelized phenomenon

For the spin wave propagating in a uniformly magnetized waveguide, depending on the propagation frequency, the spin wave beam can be concentrated in the middle of the waveguide, or it can be split into two beams to propagate at the edge of the waveguide, as shown in Fig. 2 (d). The above phenomenon was named channelized spin wave propagation.

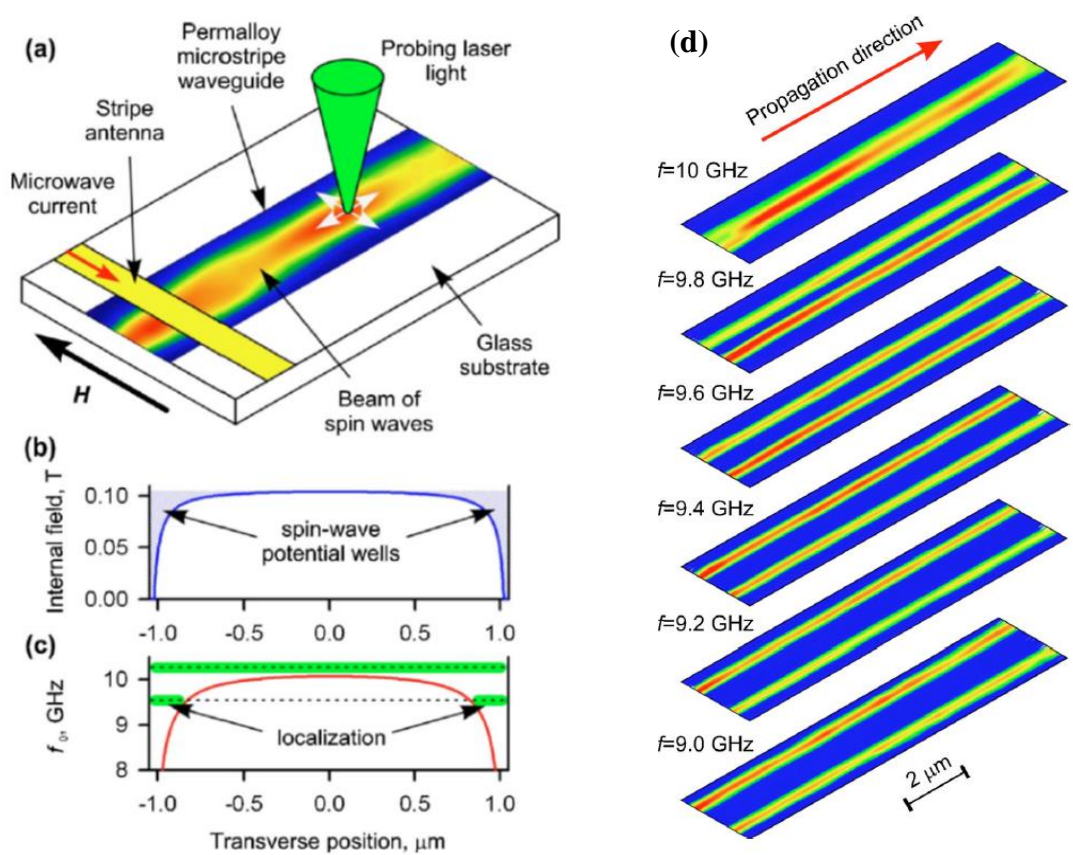

Fig. 2 (a) Schematic diagram of experimental program. (b) The distribution of the effective field along the width of the waveguide. (c) The lowest cut-off frequency distribution of the spin wave along the width of the waveguide. (d) The relationship between the beam distribution and frequency of the propagating spin wave.

This phenomenon can be explained by the demagnetizing effect in the width direction of waveguides[14]. In this experiment, the waveguide is composed of the permalloy strip with thickness of 20 nm and width of 2.1 μm. The bias magnetic field with $H_0 = 0.11T$ is applied along the direction perpendicular to the long axis of the waveguide. The Damon–Eshbach (DE) mode spin wave is excited by 1 μm wide microstrip antenna, and the propagation of spin wave is probed by Micro-focus Brillouin light scattering (μBLS) spectroscopy, as shown in Fig. 2 (a). Although the external magnetic field applied along the direction perpendicular to the long axis of the waveguide is uniform, demagnetization field is still generated in this direction due to the size limitation. The effective field distribution in this direction is shown in Fig. 2 (b) after considering the demagnetizing field. The effective field at the edge is significantly smaller than that at the center, as can be seen from Fig. 2 (b). The spin wave potential well is formed at the edge of the waveguide. Furthermore, the lowest cut-off frequency distribution of the spin wave along the width of the waveguide is calculated according to the dispersion relation, as shown in Fig. 2(c). It can be seen from Fig. 2 that the spin wave beam can be distributed over the entire waveguide width with excitation frequency $f > 10\ GHz$. The spin wave beam can only be distributed at the edge of the waveguide with excitation frequency $f < 10\ GHz$ which leads to the phenomenon of channelized spin wave propagation.

## 2.2.2 Quantization phenomenon

For magnets of limited size, wave vector of spin waves may be discretized or quantized in the direction of limited size. Perpendicular standing spin waves (PSSW) formed in the thickness direction of the magnetic film is a typical case. However, the dimensions in both the thickness direction and the width direction are limited in uniform strip waveguides. Considering the limited size in the width direction, the wavevector in this direction has the following equation:

$$k_w = \frac{n\pi}{w_{eff}}\ (n = 1,2,3\dots) \qquad (8)$$

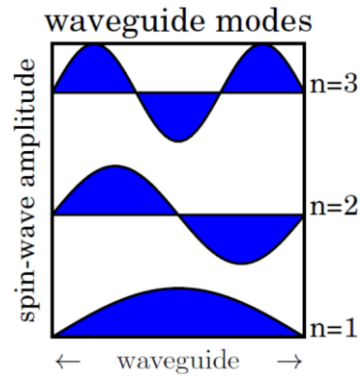

Fig. 3 Schematic diagram of width mode of spin wave (n = 1,2,3)

The spin wave propagation diagram in the uniform spin wave waveguide was probed using μBLS spectroscopy, and the experimental results showed obvious quantization phenomenon[15]. Fig. 4 shows the experimental results of the waveguide with width of 1μm and thickness of 20 μm. This experiment studies the DESW. The dotted line in Fig. 4 is the theoretical calculation curve of the dispersion relation of a large film of the same thickness. The horizontal solid line is the curve calculated using Eq. 9 and taking $k_\parallel$ as the discrete value.

$$w_{DE} = \frac{\gamma}{2\pi}[H(H + 4\pi M_s) + (2\pi M_s)^2(1 - e^{-2kd})]^{1/2} \tag{9}$$

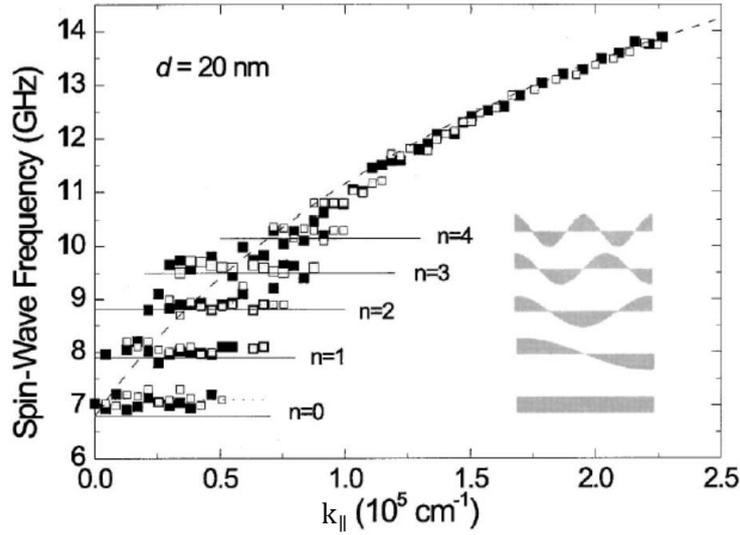

Fig. 4 Experimental results of quantization of spin wave propagation in uniform spin wave guide.

## 2.2.3 Self-focusing phenomenon

Demidov et al. utilized μBLS spectroscopy to explore the propagation characteristics of DESW in uniform waveguides[16-17]. They found that the spin wave propagates unevenly along the waveguide and has a complex relationship with the spatial coordinates. For the region near the excitation antenna, the spatial distribution of the spin wave beam is wider, and there are two maximum peaks at the edge of the region and a minimum peak at the center of the region. For the region far away from the excitation antenna, the tangential distribution of the spin wave beam changes, a maximum peak appears in the center of the strip, and the beam width decreases, as shown in Fig. 5.

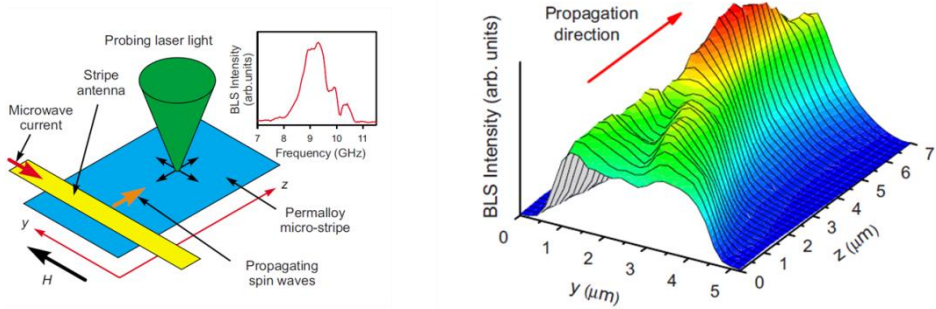

Fig. 5 Schematic diagram of the self-focusing phenomenon of the spin wave in a uniform waveguide.

The self-focusing phenomenon can be explained by the interference effect between the quantized modes in the width direction of the spin wave in uniform waveguides. As mentioned in the previous section, the wavevector component parallel to the width of the strip is quantized, while the wavevector component along the long axis of the strip is continuous. This quantization phenomenon makes the dispersion curve of DESW split into a cluster of curves, because the wave has different antinodes along the width, as shown in Fig. 6 (d). It should be noted that the even-order mode cannot be excited, only the odd-numbered mode can be excited, and the excitation efficiency is proportional to $1/n^2$ due to the uniformity of the magnetic field excited by the antenna microwave. Thus, the mode analysis for n>5 can be ignored, that is, only the relationship between n=1 and n=3 modes is considered. Fig. 6 (a) and (b) are the propagation diagrams of the spin wave mode with n=1 and n=3 when there is no interference effect, respectively. Due to the difference in phase velocity between different modes, spin waves of two modes with the same frequency will interfere when propagating along the waveguide axis at the same time. The interference period is $l = \frac{2\pi}{\Delta k}$.

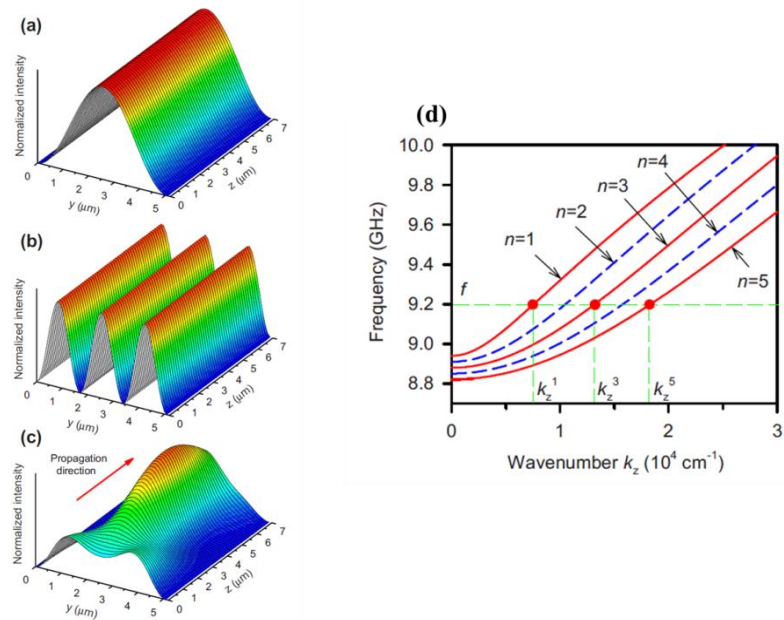

Fig.6 Schematic diagram of the propagation of spin waves with different modes (a) n = 1, (b) n = 3, (c) n = 1 and n = 3. (d) Dispersion relationship of waveguide width quantization mode.

The dynamic magnetization component of the n-order mode is as follows:

$$m_n(y,z) \propto sin(\frac{n\pi}{w}y)cos(k_z^n - \omega t + \varphi_n) \qquad (10)$$

where $k_z^n$ is the longitudinal component of the n-order spin wave mode wavevector and $\varphi_n$ is the phase of excitation source. The intensity distribution of the spin wave is proportional to the average value of $m_n(y,z)$ in the oscillation period $2\pi/\omega$. In the process of calculating the interference image, utilizing $m_1(y,z) + \frac{1}{3}m_3(y,z)$ (the excitation efficiency of spin wave modes of different orders is considered here), the total spin wave intensity is

$$I_\Sigma(y,z) \propto sin(\frac{\pi}{w}y)^2 + \frac{1}{9}sin(\frac{3\pi}{w}y)^2 + \frac{2}{3}sin(\frac{\pi}{w}y)sin(\frac{3\pi}{w}y)cos(\Delta k_z + \Delta\varphi) \qquad (11)$$

where $\Delta k_z = k_z^3 - k_z^1$ and $\Delta\varphi = \varphi_3 - \varphi_1$. It can be seen from Eq. 11 that the interference between spin waves of different modes is periodic, and the period is $l = \frac{2\pi}{\Delta k_z}$. The relative phase shift between different modes is represented by $\Delta\varphi$. Fig. 6 (c) shows the intensity of the interference spin wave, from which it can be seen that Eq. 11 is in good agreement with the experimental results. It can be seen from the above equation that the narrower width of the stripe will lead to the more obvious separation of the dispersion curve caused by the transverse quantization of spin waves, and the smaller interference period.

### 2.2.3 Nonlinear self-modulation effect of spin wave beam width

Fig. 7 is the propagation diagram of the spin wave under different excitation power, the excitation frequency is $f = 8.8\ GHz$, and the external field is $H_0 = 830\ Oe$. It can be seen from Fig. 7 that the beam width of spin wave changes nonlinearly with the increase of excitation power. This change is named transverse self-modulation[18]. The spin wave is transmitted linearly in the waveguide with the excitation power $P = 10mW$, and the energy of the spin wave is periodically concentrated in the middle of the waveguide (this feature can be explained by the self-focusing effect in the previous section). In this case, the width of the spin wave beam only slightly changes, that is, there is no significant difference in the beam width between the focused and the unfocused regions, as shown in Fig. 7 (a). With the increase of the excitation power, the beam in the focused region is compressed nonlinearly, while the beam in the unfocused region is nonlinear broadened. This makes the phenomenon of nonlinear self-modulation of spin wave beam width significant.

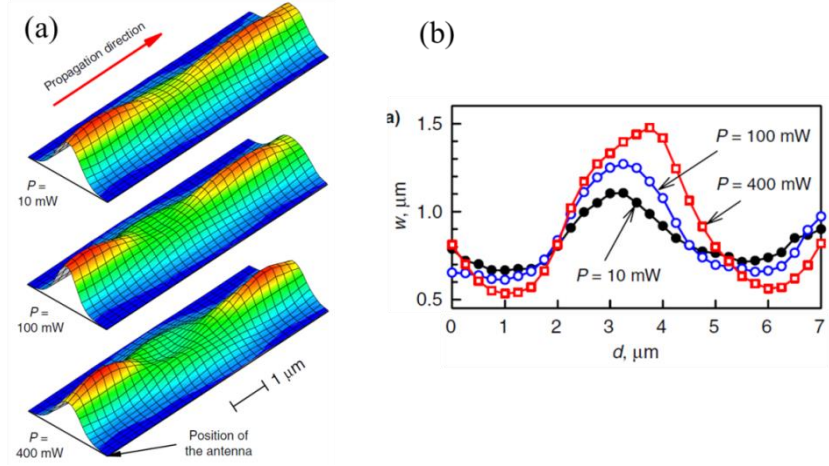

Fig. 7 (a) The distribution of spin wave intensity at different excitation frequencies. (b) The relationship between the beam width and the propagation distance at different excitation frequencies.

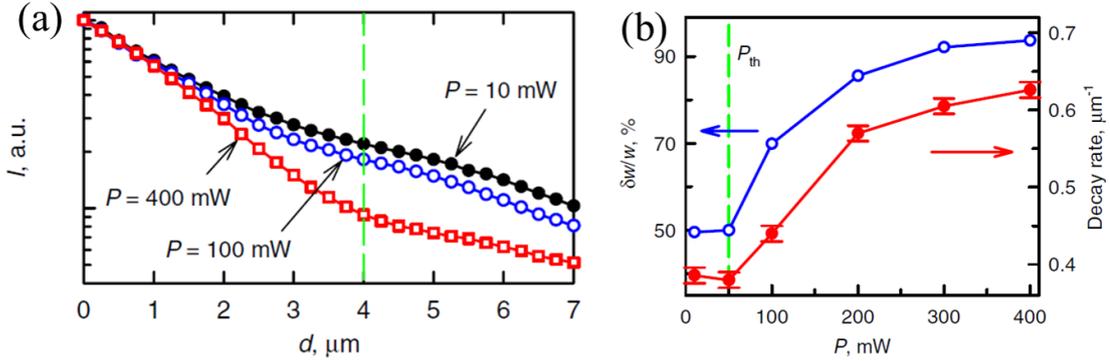

Fig. 8 (a) The relationship between the intensity of the spin wave in the width direction and the propagation distance. (b) The relationship between modulation depth, spin wave propagation attenuation rate and excitation power.

This abnormal nonlinear phenomenon is the result of the competition between the nonlinear frequency shift of various width modes of spin wave and the nonlinear damping caused by four magnon scattering. The change of damping affects the amplitude of spin wave. Obviously, the increase of damping will increase the decay rate of spin wave. Fig. 8 shows the relationship between nonlinear self-modulation effect and nonlinear damping. The self-modulation depth can be defined as:

$$\frac{\delta w}{w} = \frac{2(w_{max} - w_{min})}{(w_{max} + w_{min})} \quad (12)$$

Compared with Fig. 7 (b) and Fig. 8, it is found that the self-modulation depth increases with the increase of the decay rate of spin wave. And the self-modulation depth has the same threshold power as the decay rate of spin wave (here is 50mW).

## 2.2.4 Tunneling effect

In quantum mechanics, the tunneling effect refers to the quantum behavior in which microscopic particles such as electrons can enter or pass through high potential barriers (the height of potential barriers is greater than the total energy of particles). It is a manifestation of the volatility of quantum mechanics particles. In 1928, George Gamow proposed to use the tunneling effect to explain the alpha decay of atomic nuclei. Spin wave originates from the inherent spin of the particle. Thus, the tunneling effect can also occur in the propagation of spin waves.

The potential barrier in the uniform spin wave waveguide is composed of magnetic inhomogeneity micro-regions. These magnetic inhomogeneity micro-regions can be realized in many ways such as it is formed in the propagation direction of the waveguide by utilizing an energized conductor or generated in the region where the saturation magnetization changes. In the 1960s, the propagation of spin waves in a non-uniform magnetic field has been studied. Schlömann first noticed the similarity between the propagation of exchange mode spin waves and the motion of quantum mechanical particles[19]. Under the condition of ignoring the magnetic dipole exchange effect and magnetic anisotropy, the LLG equation describing the dynamic characteristics of magnetization can be rewritten as the static Schrodinger equation, where $m \propto exp(i\omega t)$ is similar to the wave function, and the magnetic field is used as potential energy, then[20]:

$$-\frac{2A}{M_s}\frac{\partial^2 m}{\partial z^2} + [H(z) - \frac{\omega}{\gamma}]m = 0 \qquad (13)$$

where $A$ is the exchange stiffness, $M_s$ is the saturation magnetization, $\gamma$ is the gyromagnetic ratio and $z$ is the direction of spin wave propagation. From Eq. 13, it can be concluded that the dispersion relation of plane spin wave ($m \propto exp(ikz)$) is as follows:

$$\omega = \Delta(z) + \frac{2\gamma A}{M_s}k^2 \qquad (14)$$

where $\Delta(z) = \gamma H(z)$ is the band gap of the dispersion spectrum, which is very similar to the dispersion of microscopic particles in the potential field $U(z)$:

$$E = U(z) + \frac{h^2}{2m}k^2 \qquad (15)$$

It is assumed that the spin wave with frequency $\omega$ enters the inhomogeneous magnetic region $H = H(z)$ and the band gap $\Delta(z) = \gamma H(z)$ is appropriate. In this case, by changing the wavevector of spin wave in the region $k = k(z)$, the dispersion relation of Eq. 14 can be established, and the spin wave can tunnel through the region. However, if the difference between the band gap and $\omega$ is too large, there will be no wave vector satisfying the dispersion relation at this frequency, so the spin wave will be reflected in this region. Therefore, in this case, the region becomes a potential barrier that hinders the propagation of spin waves.

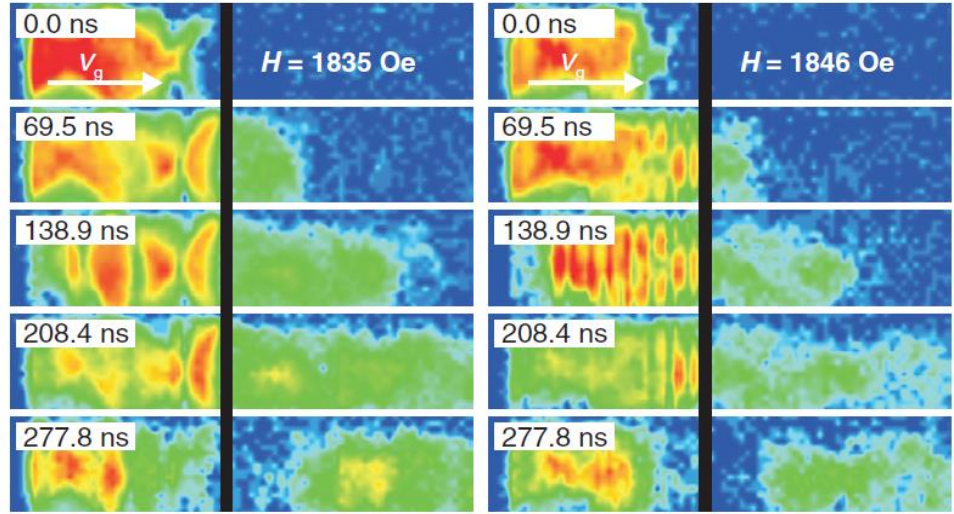

Fig. 9 Tunneling of BVMSW in different magnetic fields.

Spin waves can tunnel through the magnetic inhomogeneous micro region by the exchange effect, when the size of the magnetic inhomogeneous region is close to the exchange length. Under the condition that the size of the magnetic inhomogeneous region is much larger than the exchange length, if the wavelength of the spin wave is greater than the width of the magnetic non-uniform region, the spin wave can tunnel through this region by dipole-dipole interaction. In 2010, Schneider et al. utilized μBLS spectroscopy to observe the phenomenon of spin waves tunneling through the magnetic inhomogeneous region by the interaction of dipole-dipole interaction[21]. As shown in Fig. 9, the backward volume magnetostatic spin wave (BVMSW) configuration is used in the experiment, and the magnetic inhomogeneity region is composed of an air gap with width of 38 μm (represented by black lines in Fig. 9). Spin waves is excited at the left end of the waveguide, the excitation frequency is 7.132 GHz, and the pulse width is 200 ns. It can be seen from Fig. 9 that most of the spin waves are reflected in the air gap, and standing waves are formed on the left side of the air gap. But a small part of the spin waves tunnel through the air gap. Furthermore, it can be seen that tunneling is more likely to occur when the applied magnetic field decreases.

# 3. Modulations of spin wave propagation characteristics in uniform waveguides

The heart of a spin wave device is a waveguide that transmits and processes spin waves. The dispersion characteristics of spin wave depend on the effective field in the waveguide. Both the structure of the waveguide and the internal micromagnetic structure affect that. In addition, the effective field is also related to the dipole field between materials and external field. This section describes some typical methods of modulating spin wave propagation characteristics in uniform waveguides. We introduce it from three aspects: structure modulation, external field modulation and dipole field modulation.

## 3.1  Structure modulation methods

As described in Eq. 4 in Sec. 2, the dispersion relationship of spin waves is closely related to the effective field of the waveguide. The effective field in the waveguide can be changed by changing the microstructure or forming topological magnetic structures. This section describes the above two aspects with examples.

### 3.1.1 Microstructure modulation methods

There are many ways to change the structure of magnetic media such as changing the width of the magnetic media and setting defects. Compared with other modulation methods, this type of modulation method is more direct and stable.

Demidov et al. proposed a variable-width waveguide structure, as shown in Fig. 10 (a) [26]. Damon-Eshbach mode configuration was adopted in this experiment. As mentioned in Sec. 2, the phenomenon of channelized spin wave propagation occurs due to the existence of the spin wave potential well in uniform waveguides. The spin wave potential well is closely related to the demagnetization effect in uniform waveguides. And the demagnetization effect is different in different width waveguides. As shown in Fig. 10 (b), the internal magnetic field in the wider waveguide is more uniform, while there is obvious inhomogeneity in the narrower waveguide. The difference in the internal magnetic field distribution makes the dispersion curve different. Fig. 10 (c) depicts the dispersion curves at the wider and narrower waveguides. The dispersion curve at the narrower waveguide shifts down about 0.5 GHz compared with that at the wider waveguide. This shift causes the special propagation phenomenon of spin wave shown in Fig. 10 (d). Fig. 10 (d) shows that as the width of the waveguide increases, the spin wave propagating in the narrow part of the waveguide exhibits the phenomenon of channelized. Utilizing this kind of waveguide structure, the mode conversion of spin wave can be realized effectively (unchannelized transmission to channelized transmission). This type waveguide structure can be used for spin wave interference to design spin wave logic devices. For example, we can build a narrow-wide-narrow waveguide and change the phase of the spin wave on one side of the wide waveguide in other ways, thus forming the function of the spin wave inverter.

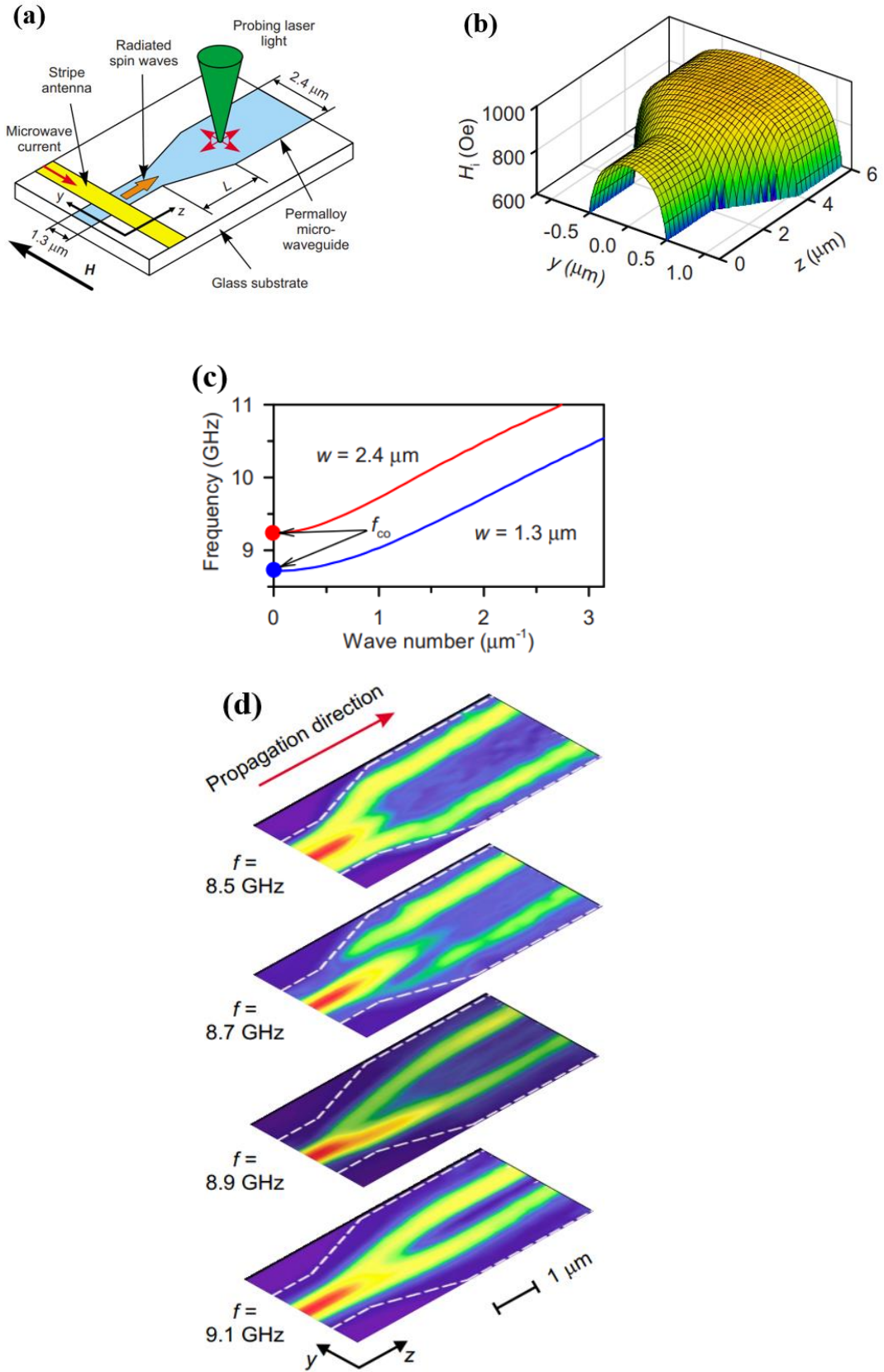

Fig. 10 (a) Diagram of the experiment. (b) The distribution of the internal magnetic field in the waveguide. (c) The dispersion curves at the wider and narrower waveguides. (d) Diagram of intensity distribution of spin wave at different excitation frequencies.

Daniel et al. studied the diffraction of spin waves propagating in Permalloy waveguide with a submicron-sized circular defect[27]. In this experiment, the transmission characteristics of DESW are studied. The existence of defects leads to the local inhomogeneity of the internal magnetic field. This local inhomogeneity leads to the change of spin waves propagation characteristics in this region. It can be seen from Fig. 11 that the complexity of the spin wave intensity mode increases dramatically when the spin wave encounters the defect. The complex diffraction pattern after the defect can be considered as the interference of several modes in the waveguide. As shown in Fig. 11 (b), the spin wave is converted from the fundamental mode to the higher-order mode after passing through the defect due to the spin wave interference caused by the defect. The work of Daniel et al. showed that the defect can cause local changes in the internal magnetic field of the waveguide, thereby changing the propagation characteristics of the spin wave. It can be seen from Fig. 11 (b) that defects can not only change the mode of the spin wave but also change the propagation direction of the spin wave. In theory, reasonable use of defects in the design of spin wave devices can build spin wave logic devices with any logic function. Recently, Qi Wang confirmed this view in his published papers[28]. Qi Wang realized the method of inversely designing spin wave devices by using defects, that is, first specifying the function of the spin wave device, and then determining the position and the number of defects through an algorithm, as shown in Fig 12. Therefore, it may be feasible to design spin wave devices by making use of the size, number and position of defects to make the spin wave propagation characteristics change locally.

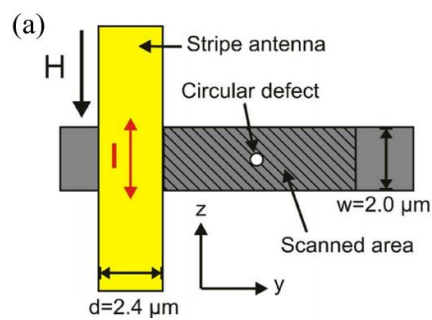

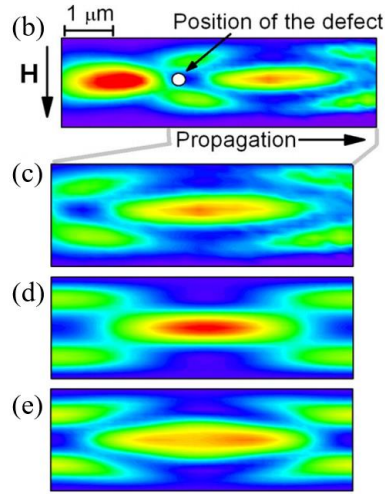

Fig. 11 (a) Diagram of the experiment. (b) The intensity graph of the spin wave with an excitation frequency of 8.9 GHz. (c) The spin wave intensity graph of the part after the defect. (d) The spin wave intensity diagram calculated by n = 1 and n = 3 modes. (e) The spin wave intensity diagram calculated by n = 1, n = 3 and n = 5 modes.

(a)

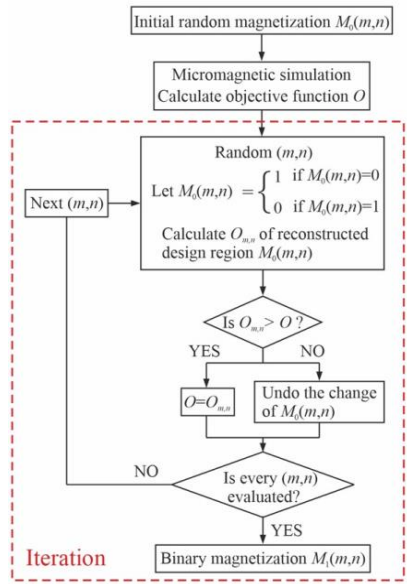

(b)

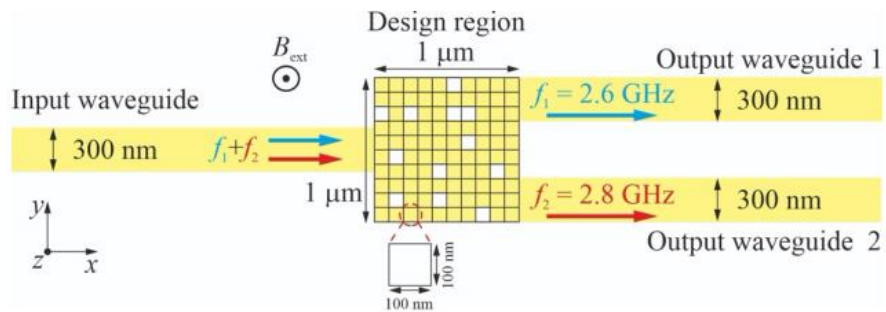

Fig. 12 (a). Flow chart of the Direct Binary Search (DBS) algorithm for one iteration. (b)

Structure of the inverse-designed magnonic frequency demultiplexer.

This section introduces two methods of modulating spin waves with microstructure. Both methods directly change the structural parameters of the waveguide. The channelization of spin waves can be controlled by changing the width of the waveguide. The defects in the waveguide can be used to realize arbitrary logic spin wave logic devices. This type of method is suitable for the design of fixed function devices. In addition, this type method can be combined with the modulation methods described in the following sections to design reconfigurable spin wave logic devices. However, such methods have strict requirements for process level.

## 3.1.2 Micromagnetic structure modulation methods

The internal non-uniform effective field of micromagnetic structures (topological magnetic structures) such as domain walls and vortexes affect the propagation characteristics of spin waves. The advantage of using topological magnetic structures to modulate spin waves is that the generation and annihilation of magnetic topology can be controlled by other modulation methods to achieve the purpose of dynamic control. In recent years, the research on topological magnetic structures has attracted the attention of more and more researchers.

In magnetic materials with perpendicular magnetic anisotropy, domain walls (DW) interact with spin waves. DW affect the amplitude and phase of spin wave, and spin waves move DW through spin torque effect. Wojewoda et al. studied the propagation characteristics of spin waves with different frequencies after passing through a Néel domain wall[29]. This experiment utilizes the structure shown in Fig. 13 to study the propagation characteristics of DESW through magnetic domains. By intensity and phase resolved μBLS experiments, they found that the spin wave with an excitation frequency of 7.15 GHz separate into two beams after passing through DW (which is similar to the channelization phenomenon mentioned above), and the spin wave with an excitation frequency of 9.00 GHz be confined to the center of the waveguide after passing through the magnetic domain, as shown in Fig. 14. In addition, through wavelength fitting and comparison with the single domain state, the spin wave with an excitation frequency of 7.15 GHz has the phase shift of about $0.6\pi$ through the magnetic domain, and the phase shift of 9.00 GHz is about $0.5\pi$, as shown in Fig. 14(e) and (f). This experiment revealed the influence of DW on the propagation characteristics of spin waves. DW affects the phase and propagation characteristics of spin waves. Therefore, the utilize of external fields to realize the generation and annihilation of DW can realize controllable spin wave phase shifts and further realize spin wave logic devices. Utilizing DW can also make structures such as spin wave majority gates simpler. However, domain walls are highly sensitive to temperature, and it is difficult to control the phase shift accurately. Therefore, the implementation of spin wave logic devices based on DW still needs to overcome major difficulties.

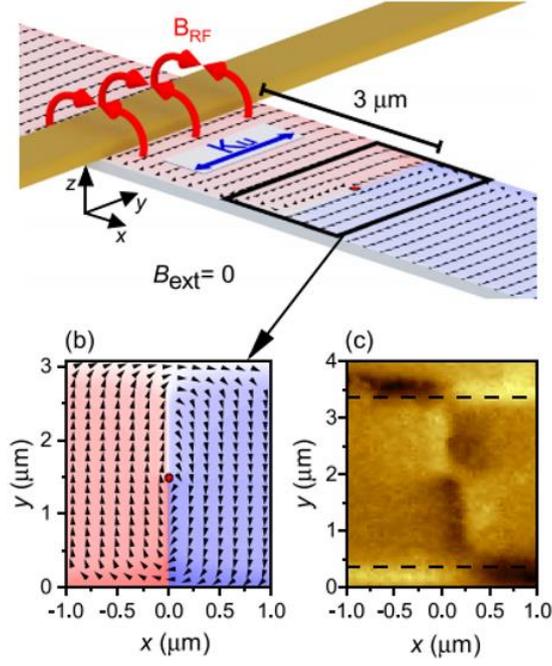

Fig. 13 Scheme of the experimental geometry.

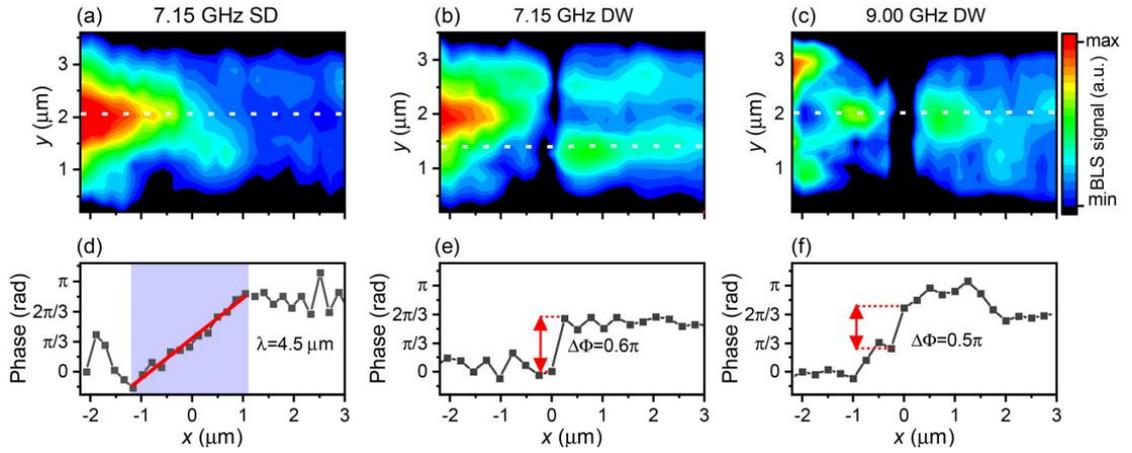

Fig. 14 (a) Intensity diagram of a spin wave with the excitation frequency of 7.15GHz in the single domain state. (b) Intensity diagram of the spin wave with an excitation frequency of 7.15GHz in the two-domain state. (c) Intensity diagram of the spin wave with an excitation frequency of 9.00GHz in the two-domain state. (d) – (f) are the phase diagrams of (a) – (c) respectively.

Vortex is also a typical topological magnetic structure. In the non-core region of vortex, the magnetization revolves around the center, while in the core region, the magnetization direction is out of plane. The overall magnetization follows the chirality principle. Park et al. studied the propagation characteristics of spin waves after passing through a vortex[30]. In this simulation, the vortex structure is realized by using the tow-domain state waveguide, as shown in Fig. 15. Fig. 16 shows the propagation diagram of spin waves through vortex with low (0.18 ~ 3.1 GHz) and high (6.4 ~ 11.6 GHz) frequencies, respectively. It can be seen from Fig. 16 (a) that the spin wave is well localized in the domain wall. The spin wave with excitation frequency of 1.8 GHz interacts with the vortex strongly. As the frequency increases, the spin waves scattered to other orthogonal arms gradually decrease. Fig. 16 (b) describes the spin wave propagation diagram of the other three orthogonal arms at different frequencies. It can be seen that the amplitude and phase of the spin wave change obviously after passing through vortex, which is related to the frequency. This work introduces a new method of spin wave phase shift. Utilizing this method and setting an appropriate structure can realize the spin wave majority gate. However, the amplitude of the spin wave is significantly attenuated after passing through the vortex, which is not conducive to the detection of the spin wave and the setting of the threshold of the spin wave logic device.

(a)

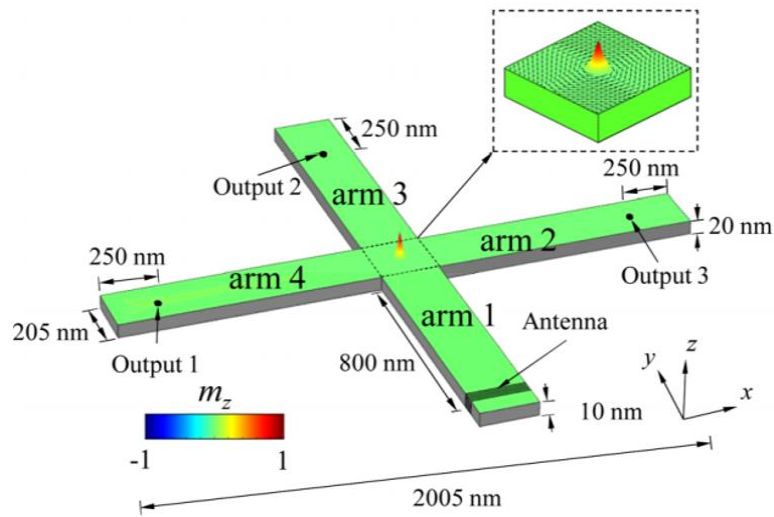

(b)

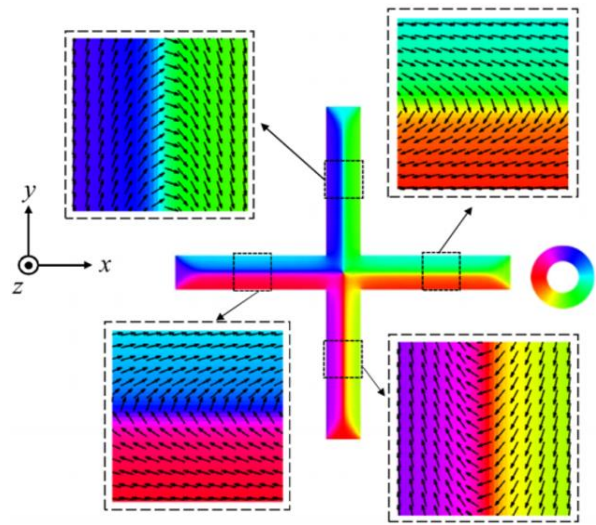

Fig. 15 (a) Scheme of the simulation geometry. (b) Initial magnetization state of structure in (a).

(a)

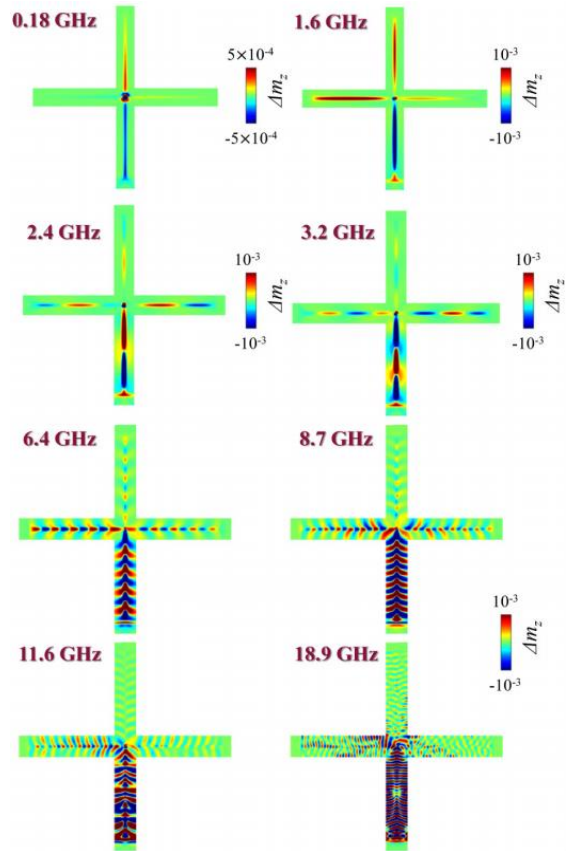

(b)

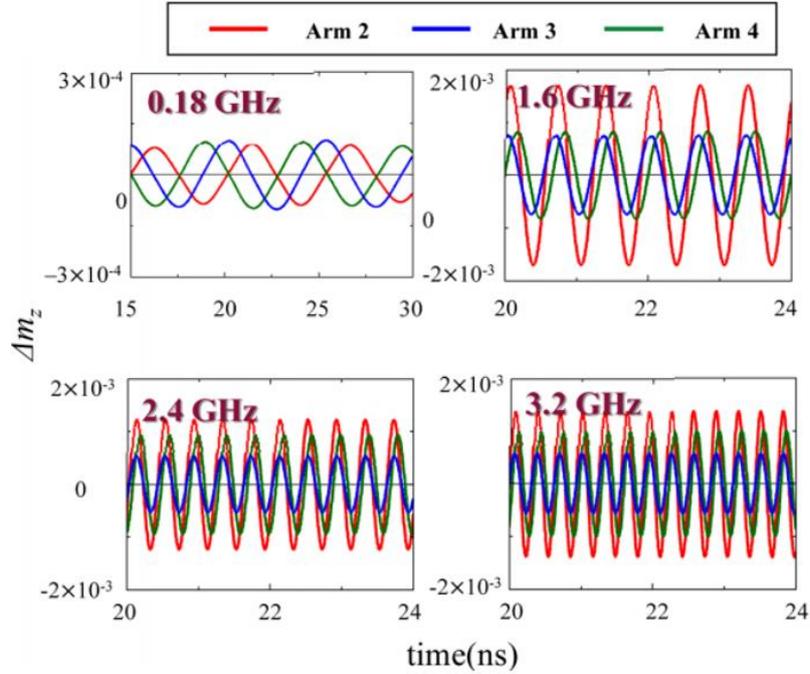

Fig. 16 (a) Diagrams of spin wave propagation with different excitation frequencies. (b) The spin wave propagation diagram of the other three orthogonal arms at different frequencies.

This section introduces two methods of modulating spin waves using topological magnetic structures. The topological magnetic structures have influence on the mode, phase and amplitude of spin waves. In addition, this kind of modulation method can be combined with the external field to realize the generation and annihilation of topological magnetic structures, and realize the reconfigurable spin wave logic device. However, spin waves have a large attenuation through these topological magnetic structures. How to reduce the loss of spin waves is the biggest challenge in the application of this kind of methods.

## 3.2. Modulation methods of external stimulus source

The magnetic properties of waveguides are affected by many kinds of stimulus source such as STT/SOT, dipolar-coupling, magnetic fields, electric fields and stress fields. Applying an external stimulus source to waveguides is a direct and efficient way to change the effective field in the uniform waveguides. This section will introduce several ways of using external stimulus source to modulate the propagation characteristics of spin waves.

### 3.2.1 Modulation method of utilizing STT or SOT effects

Spin-polarized current can interact with magnetic moments and transfer its spin angular momentum to the magnetic moment to realize the reversal of the magnetic moment. This phenomenon is called spin-transfer torque effect (STT). Another effect similar to its effect is called spin orbit torque effect (SOT). The difference between STT and SOT lies in the way the spin current is generated. The influence of STT and SOT on magnetization dynamics can be explained by the following equation[31]:

$$\frac{\partial m}{\partial t} = -\gamma \boldsymbol{m} \times H_{eff} + \alpha \boldsymbol{m} \times \frac{\partial m}{\partial t} + \frac{\gamma}{M_s} \times \boldsymbol{\tau} \qquad (16)$$

where $M_s$ is the saturation magnetization, $\boldsymbol{m} = \boldsymbol{M}/M_s$ is the magnetization unit vector, $\alpha$ is the gyromagnetic ratio, $H_{eff}$ is the effective field, $\boldsymbol{\tau}$ is the torque term by STT or SOT. It can be seen from Eq. 16 that STT or SOT exerts an additional torque on the LLG equation. In recent years, STT and SOT have been used in magnetic memory due to their fast reversal of magnetic moment characteristics. STT and SOT effects also affect the propagation characteristics of spin waves. STT and SOT are more used to dynamically change the propagation characteristics of spin waves, that is, to design reconfigurable spin wave logic devices.

Most modulation methods utilizing STT adopt the sandwich structure. The sandwich structure is usually composed of pinning layer, space layer and waveguide. Qi Wang et al. proposed a method of controlling the spin wave spectra dynamically in a waveguide based on STT[32]. The simulation structure is shown in the Fig. 17. The comparison between Fig. 18 (a) and (c) shows that STT produces a periodic magnetic structure. This periodic magnetic structure can hinder the transmission of spin wave in a certain frequency band, that is, a forbidden band. Therefore, the generation and annihilation of this magnetic structure can be controlled by STT to control the spin wave transmission in a specific frequency band, as shown in Fig. 19. The greatest advantage of this modulation method is that the STT effect can be used for dynamic modulation, which is conducive to the frequency bandwidth expansion and reconfigurability of the spin wave device based on this design. However, the power consumption of the STT effect and the influence of heat on the saturation magnetization of materials are still great challenges for the application of this modulation method. In addition, in recent years, researchers have found that sound wave can be used to assist STT modulation, so as to reduce the power consumption of STT modulation[33].

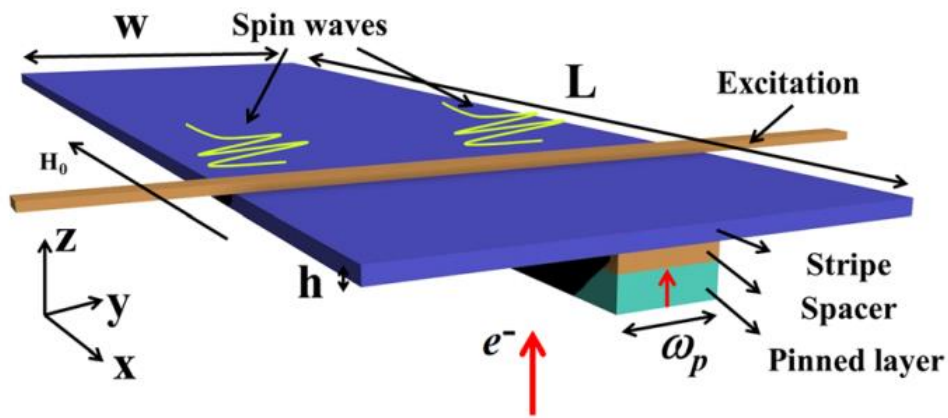

Fig. 17 The structure of spin wave is adjusted by STT based on sandwich structure.

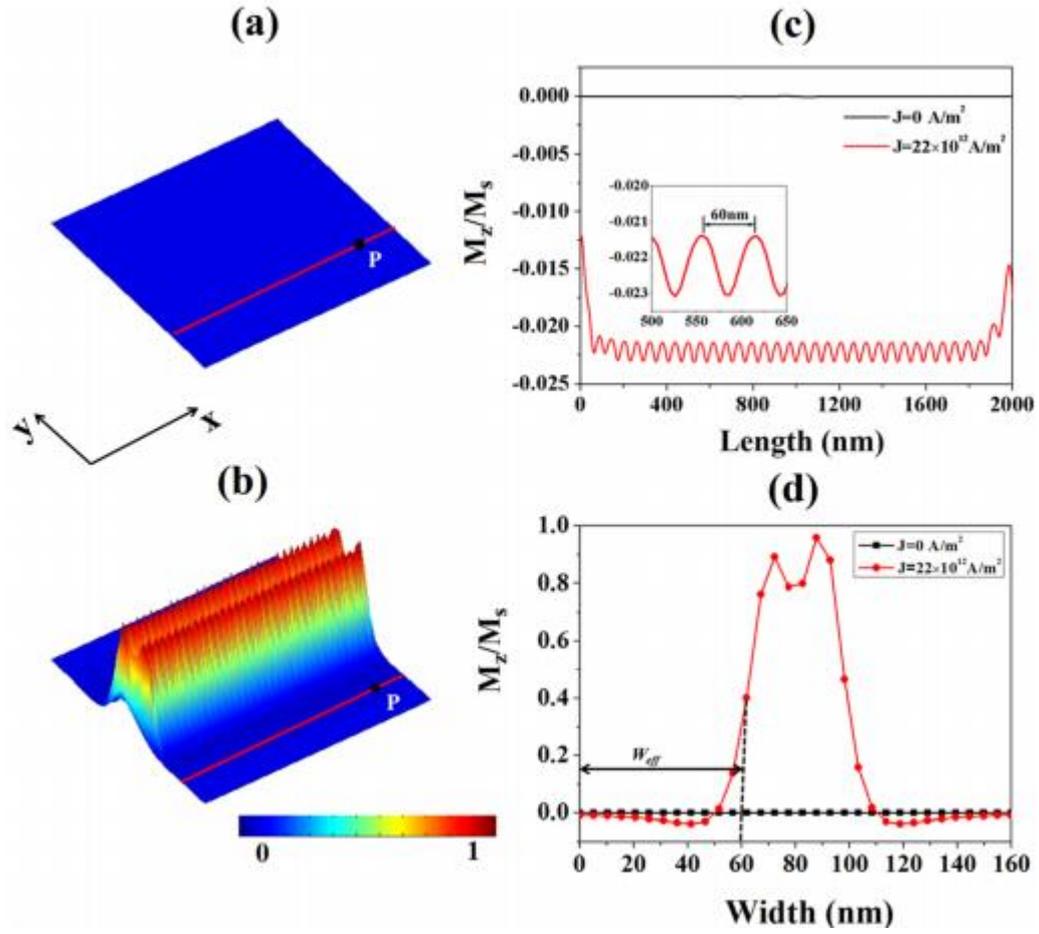

Fig. 18 (a) and (b) The magnetization (Mz/Ms) of waveguide for (a) $J = 0\ A/m^2$ (b) $J = 22 \times 10^{12}\ A/m^2$, respectively. The Mz is extracted along (c) x-axis and (d) y-axis, respectively.

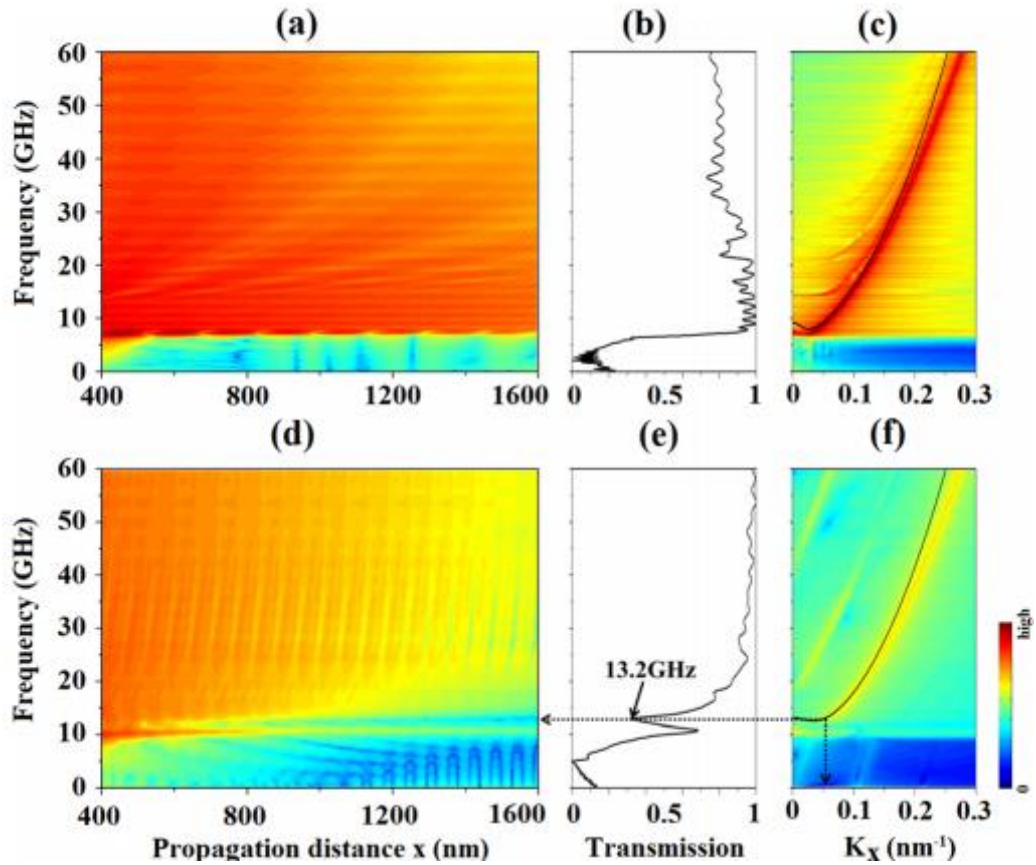

Fig. 19 Propagation characteristic diagram of spin wave without spin-polarized current (top row, plots (a) to (c)) and with spin-polarized current $J = 22 \times 10^{12}\ A/m^2$ (bottom row, plots (d) to (f)).

As a novel effect, the SOT effect has received great attention from researchers in recent years. Compared with the STT effect, the SOT effect has lower power consumption. In addition, the STT effect is an interfacial effect, which is more conducive to the miniaturization of the device design. Evelt et al. proposed an efficient modulation method to control spin wave propagation in ultra-thin YIG based on SOT[34]. The experimental structure diagram is shown in Fig. 20. The current flowing into Pt is converted into pure spin current and injected into YIG waveguide due to the spin Hall effect (SHE) effect. The injection of pure spin current results in the SOT effect. The experimental results show that the effective damping of the waveguide has a linear relationship with the current applied by the SOT modulation, as shown in Fig. 21. The amplitude of spin waves can be amplified in a certain current range. For a long time, how to propagate spin waves over long distances has also been a difficulty in the study of spin wave devices. Although there have been ideas such as spin-wave repeaters before this, those methods have not been able to efficiently realize the long-distance transmission of spin waves. It may be possible to realize a spin wave amplifier by applying the SOT modulation method. However, the amplification of spin waves is hindered by Joule heat and the excitation of large-amplitude auto-oscillations. Thus, how to overcome these two problems is the key to the application of this modulation method in practice.

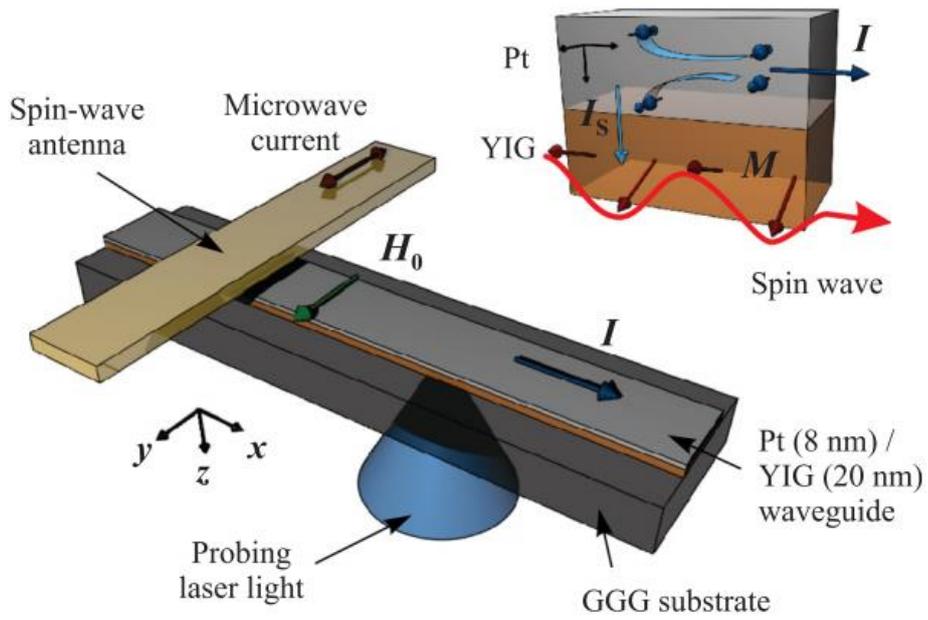

Fig. 20 Schematic of the SOT modulation implementation.

(a)

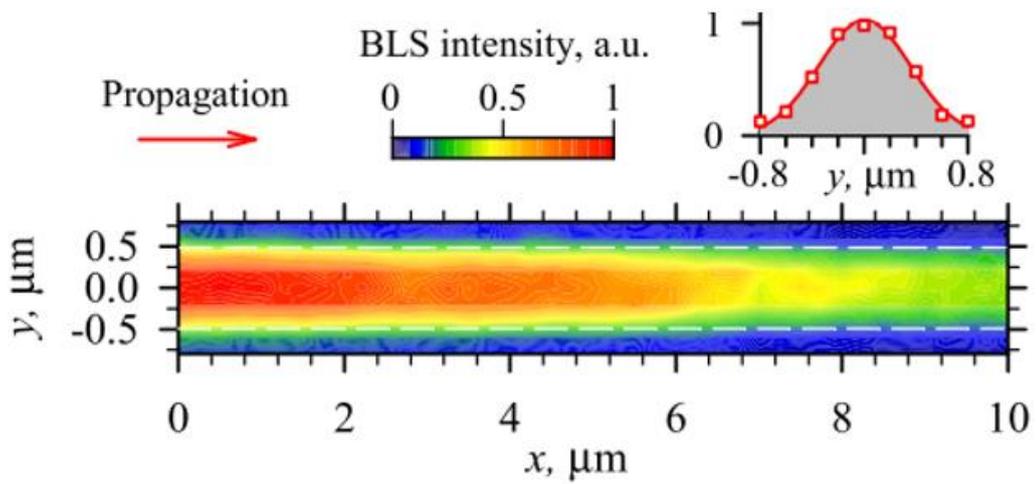

(b)

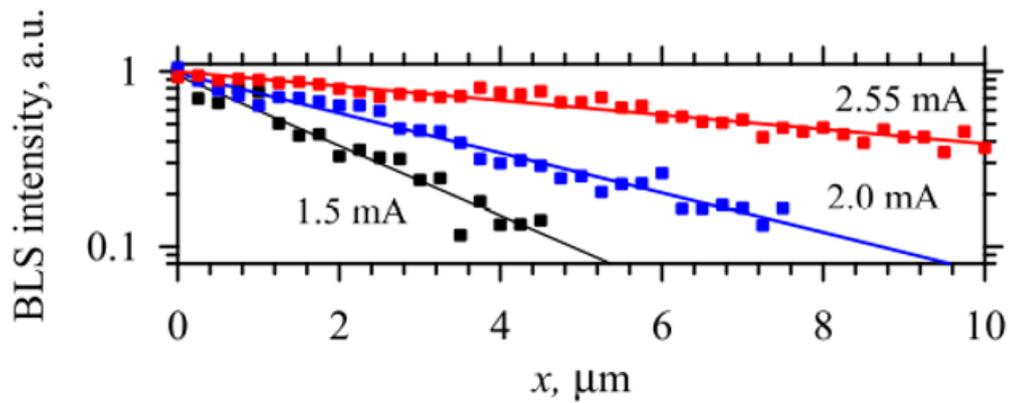

(c)

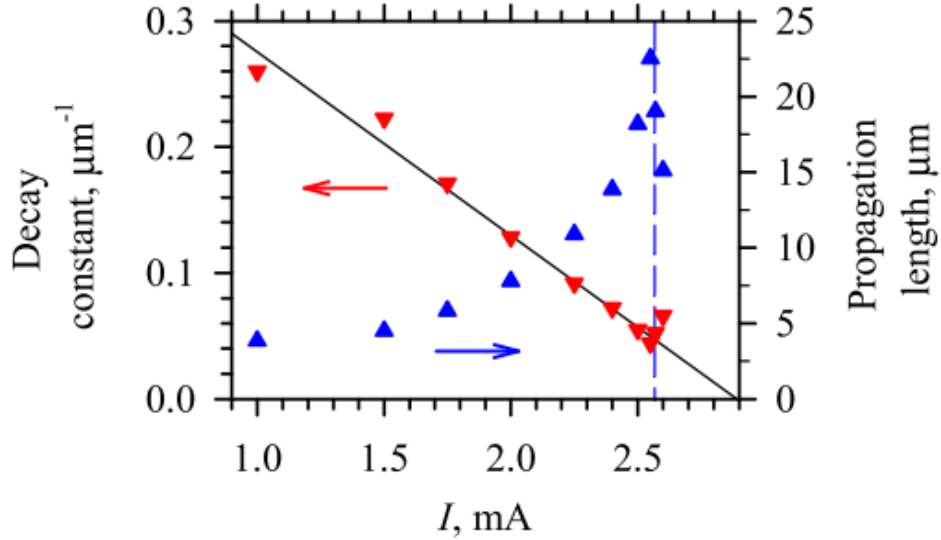

Fig. 21 (a) Spin wave intensity map with I = 2.55 mA. (b) Spin wave intensity graph with different currents. (c) The relationship between the decay constant of the spin wave and the current.

This section describes that the STT and the SOT modulations can effectively hinder or amplify the transmission of spin waves. The advantage of the STT and SOT modulations lie in their fast response speed. Although both the STT and the SOT modulations have the problem of Joule heating, the Joule heating can be reduced by reducing the width of the waveguide or integrating it into the heat-sink devices. In addition, the STT and SOT effects can effectively change the magnetization direction of nanomagnets. At present, there have been studies on the combination of this effect with dipole interaction and spin wave coupled to realize spin wave devices.

### 3.2.2 Modulation method of applying voltage or strain field

The effect of the aforementioned STT or SOT modulation method is to apply an additional torque to the magnetization dynamics. Utilizing voltage or stress field modulation is achieved by inducing anisotropic changes in magnetic materials.

The voltage modulates the propagation characteristics of spin wave by voltage-controlled magnetic anisotropy (VCMA). It is more efficient to change the magnetization dynamics than spin polarized current because it avoids the generation of Joule heat. Therefore, the voltage modulation method is expected to become a fast and energy-saving magnetization dynamics modulation method in the future. Rana et al. proposed reconfigurable nanochannels controlled by VCMA in nano-waveguides[35]. The VCMA effect can locally change the magnetic anisotropy of the waveguide, thereby changing the dispersion relationship of the spin wave locally. As shown in Fig. 22 (b), the positive voltage causes the spin wave dispersion curvea of the waveguide with the dimensions of 2 μm (length) × 200 nm (width) × 1.3 nm (thickness) to shift downward. Therefore, spin waves with frequency below the cutoff frequency can propagate locally by changing the magnetic anisotropy. As shown in Fig. 22 (c), the magnetic anisotropy of the waveguide under the Au electrode changes locally, and the spin wave below the cut-off frequency is propagated locally in the nanochannel. The VCMA effect can make the dispersion curve shift. Therefore, the VCMA effect can not only localize the spin waves of certain frequency band in the nanochannel, but also locally change the phase of the spin waves, as shown in Fig. 22 (d). This modulation method provides a new idea for modulating spin wave. Compared with STT and SOT, using the VCMA effect to modulate spin wave is more efficient. However, this modulation method requires the thickness of the waveguide to be ultrathin (less than 10 nm). As far as current technology is concerned, it is difficult to achieve a uniform waveguide with such a low thickness.

(a)

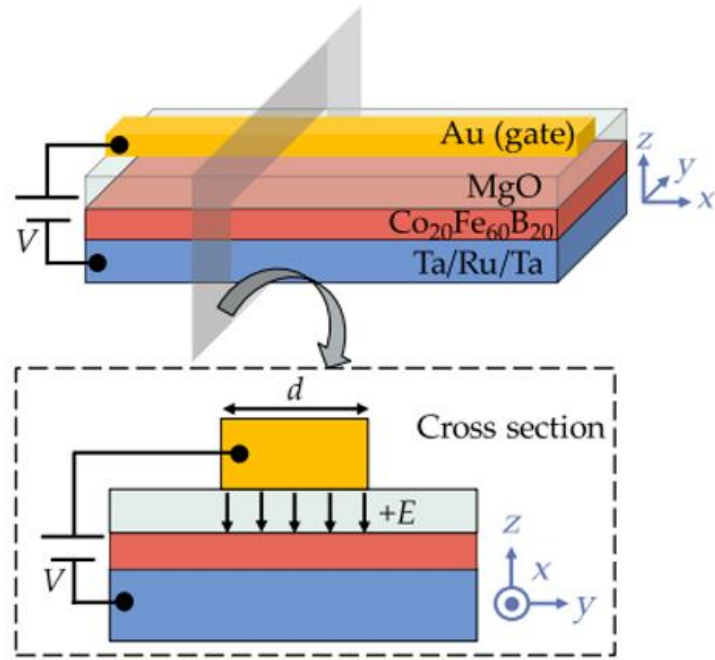

(b)

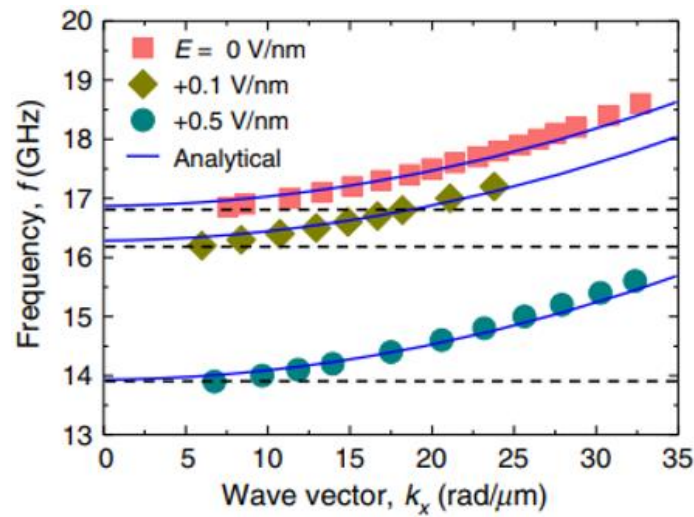

(c)

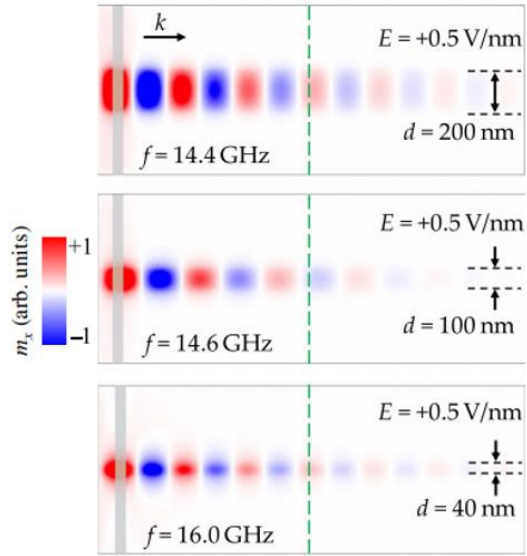

(d)

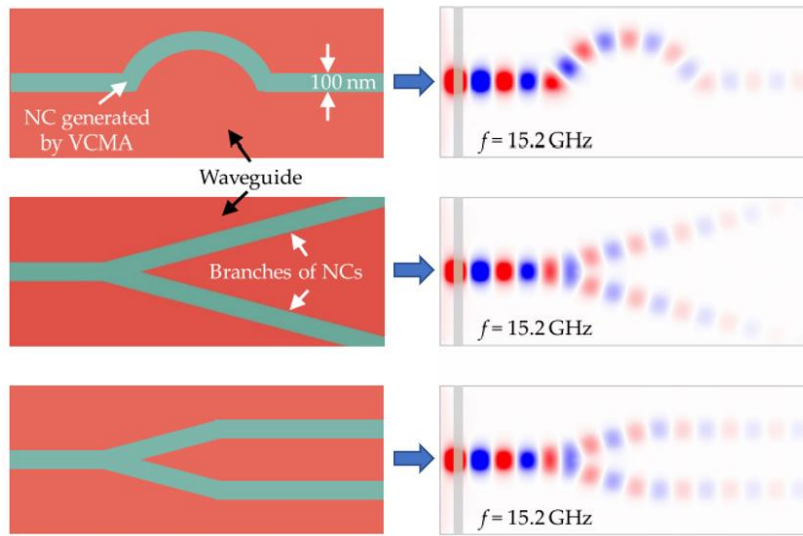

Fig. 22 (a) The structure of spin wave device based on VCMA modulation. (b) The dependence of voltage on spin wave dispersion curve of the waveguide with the dimensions of 2 μm (length) × 200 nm (width) × 1.3 nm (thickness). (c) The spin wave propagation diagram with the excitation frequency below the cut-off frequency is obtained when E = 0.5 V. (d) The shapes of different electrodes and the corresponding spin wave propagation diagram.

(a)

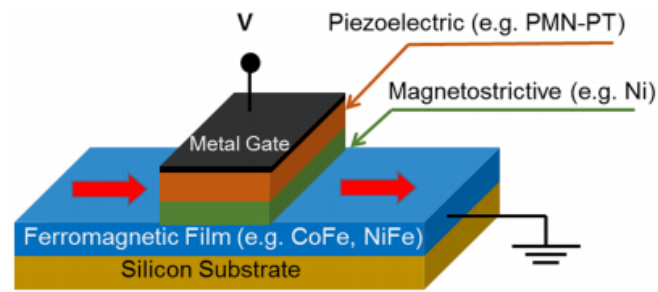

(b)

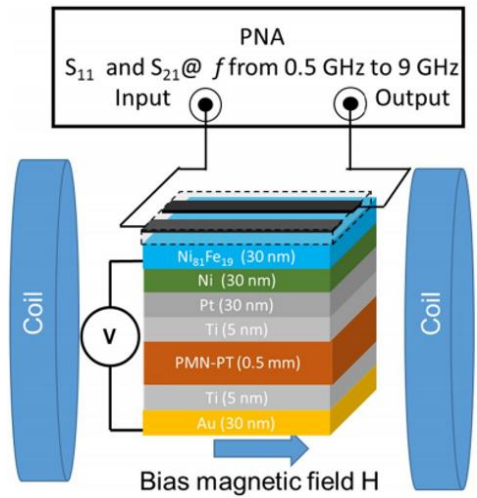

(c)

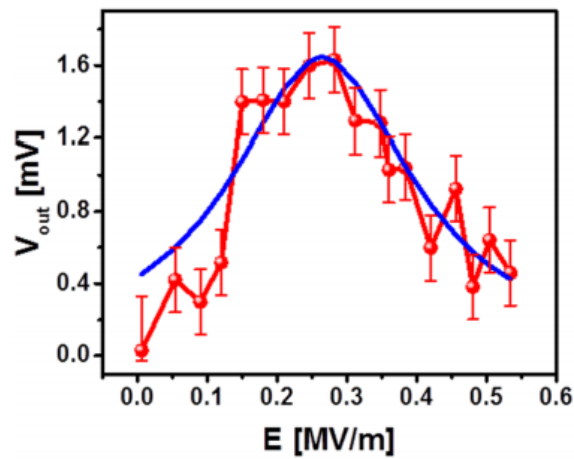

Fig. 23 (a) Schematic diagram of the stress field modulation method. (b) The experimental structure diagram of the spin wave modulated by the strain field modulation. (c) The output characteristic diagram of the spin wave modulator with the bias magnetic felid of 390Oe and the excitation frequency $f$ = 6.0GHz. The red curve shows the inductive voltage in mV produced by the propagating spin-waves. The blue curve is the result of Lorentz fitting.

The strain field modulation is mainly realized by utilizing piezoelectric (PZT) materials. This modulation structure is generally a multiferroic material composed of piezoelectric materials and magnetostrictive films. Balinskiy et al. proposed a spin wave modulator based on strain field[36], as shown in Fig. 23(b). The structure consists of the semiconductor substrate, the ferromagnetic layer, the magnetostrictive layer and the piezoelectric layer, as shown in Fig. 23(a). Spin waves propagate in the ferromagnetic layer. The strain is generated by applying an electric field to the piezoelectric material. The strain acts on the magnetostrictive layer and changes the direction of the easy axis in the magnetostrictive layer. The magnetization of the magnetic strain layer affects the magnetization of the ferromagnetic layer through exchange and dipole-dipole interaction. Finally, due to the change of the magnetization of the ferromagnetic layer, the propagation of spin waves in the ferromagnetic layer changes. The induced voltage is proportional to the amplitude of spin wave. The magnitude of the stress is proportional to the applied voltage. As shown in Fig. 23(c), the amplitude of the spin wave and the amplitude of the stress are in a quadratic function relationship. The depth of spin wave modulation is affected by the magnitude of the magnetoelastic field, that is, the magnitude of the voltage. The magnetoelastic field $H_E$ is as follows:

$$H_E = -\frac{1}{\mu_0 M_s} \begin{pmatrix} 2B_1\varepsilon_{xx}m_x + B_2(\varepsilon_{xy}m_y + \varepsilon_{xz}m_z) \\ 2B_1\varepsilon_{yy}m_y + B_2(\varepsilon_{xy}m_x + \varepsilon_{yz}m_z) \\ 2B_1\varepsilon_{zz}m_z + B_2(\varepsilon_{xz}m_x + \varepsilon_{yz}m_y) \end{pmatrix} \quad (17)$$

where $B_1$ and $B_2$ are the magnetoelastic coupled constants, $\varepsilon$ is the strain tensor and $m$ is the unit magnetization vector. Although the strain field can change the propagation characteristics of spin waves, this method involves conversion of multiple energies, and the issue of energy conversion rate needs to be considered. Despite the energy consumption problem, the discovery of giant magnetostrictive materials such as Terfenol-D makes the application of this modulation method possible.

This section introduces the propagation characteristics of spin wave modulated by VCMA and strain field. Compared with STT and SOT modulations, these two modulations do not generate Joule heat. However, these two modulations have higher requirements for material parameters. The development of suitable materials is the key to the realization of this type modulation.

## 3.3 Modulation method of dipole field

Utilizing the dipole field to modulate the propagation characteristics of spin waves is usually achieved by the dipole interaction between the high saturation magnetization material and the waveguide. The magnitude of dipole interaction $E_d$ can be described by the following formula[37]:

$$E_d \propto \frac{\mu_0}{4\pi R^3}(cos\varphi_2 sin\theta_2 cos\varphi_1 sin\theta_1 - 2cos\theta_2 cos\theta_1 + sin\varphi_2 sin\theta_2 sin\varphi_1 sin\theta_1) \quad (18)$$

where $\mu_0$ is the permeability of vacuum, $R$ is the distance between two magnetic magnets, $\theta_1$ and $\theta_2$ are the out-of-plane angles of two magnets, respectively, $\varphi_1$ and $\varphi_2$ are the in-plane angle, that is, the angle between the magnetization direction and the x-axis. Therefore, it can be seen from Eq. 18 that the dipole field can be controlled by changing the parameters such as the distance between the material and the waveguide and the magnetization direction. The final result of the change of the dipole field is the change of the effective field, so that the propagation characteristics of the spin wave change.

Y. Au et al. proposed a spin wave phase-shifting method by placing the resonator vertically[38], as shown in Fig. 24 (a). The resonators (nanomagnet) placed around the waveguide resonate when spin waves propagate in the waveguide. This resonance can produce a constructive or destructive reaction to the spin waves in the waveguide. In particular, the intensity of this reaction is related to the magnetization direction of the resonators. As shown in Fig. 24, the spin wave phase shifts in the waveguide when the magnetization direction of the resonator is in-plane. And the resonator absorbs the energy of the spin wave with out of plane magnetized, which hinders the transmission of the spin wave. This work provided a new idea for the design of spin wave logic devices. This method modulates the characteristics of spin wave without direct manipulation of the waveguide, thus avoiding to destructively change parameters of the waveguide, such as the increase of damping. If the magnetization state of the resonator can be tuned efficiently and at a high speed, it may be possible to realize a high-speed spin wave logic device or a switching device.

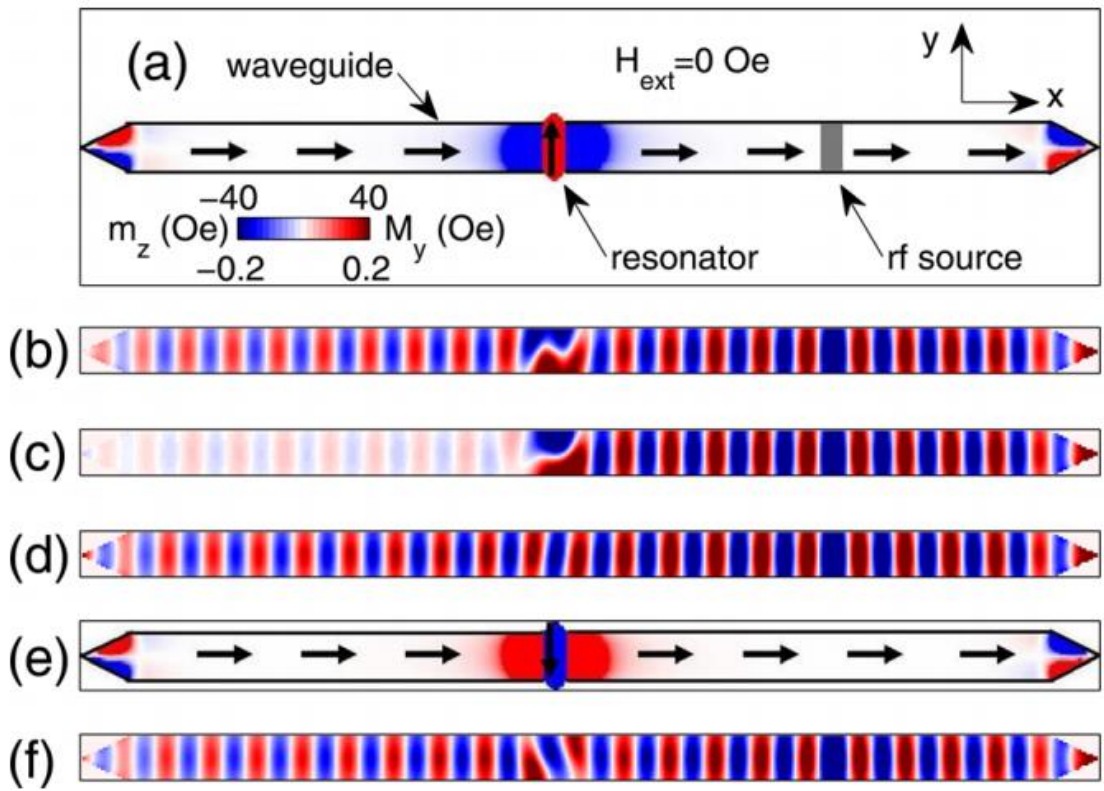

Fig. 24 (a) Structure of phase shifter based on resonator. (b) – (d) Out of plane magnetization (mz) inside the waveguide at the same relative simulation time for a vertical spacing between the resonator and the waveguide kept at 5 nm and changed to 20 and 50 nm, respectively. (e) – (f) The propagation diagram of the spin wave in the waveguide with the magnetization of the resonator toggled to negative y direction.

Similar to the previous modulation method, Zhang et al. proposed to modulate the characteristics of spin wave using dipole field generated by the material with high saturation magnetization[39], as shown in Fig. 25. In this experiment, the dipole field between Py and YIG affect the effective field on the side of YIG closer to Py. The increase in the effective field on this side increases the wavelength of the edge spin wave propagating on this side, while the spin wave on the other side is hardly affected, as shown in Fig. 25(b). The transverse placement of high saturation magnetization materials makes the phase difference and wavelength difference of spin waves which originally propagate in the same phase and the same wavelength on both sides, as shown in Fig. 25(c). In general, the dipole interaction between two nanomagnets is related to the distance between the two nanomagnets, the angle between the magnetization directions, the saturation magnetization and other factors. It is very difficult to change the structure parameters such as spacing in the working process of the device. Therefore, if the magnetization direction angle or saturation strength between two nanomagnets can be changed efficiently, the corresponding spin wave logic device function can be realized.

(a)

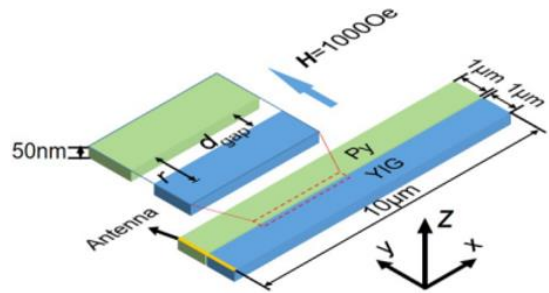

(b)

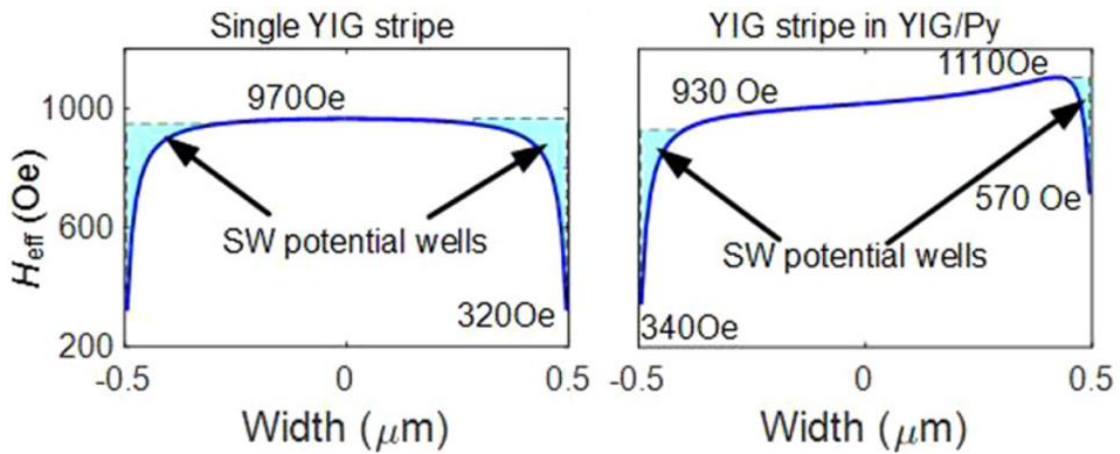

(c)

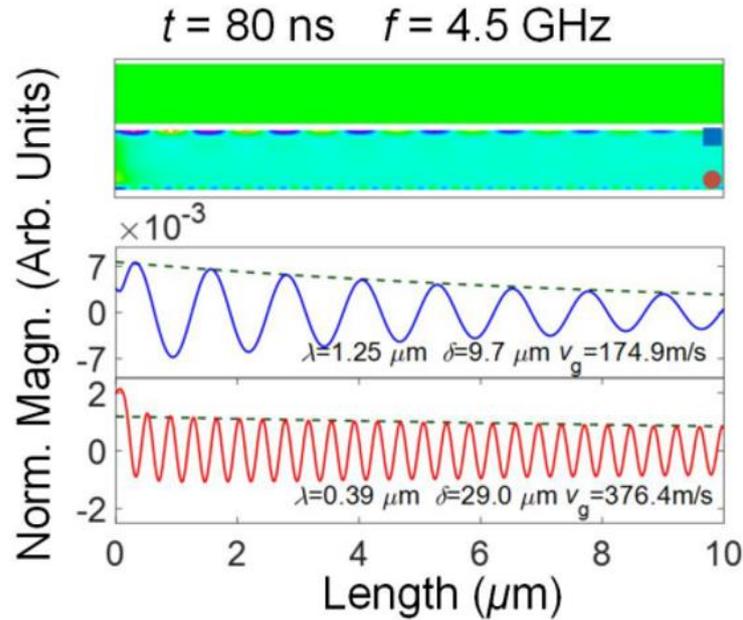

Fig. 25 (a) Structure diagram of spin wave phase shift based on dipole action. (b) The effective field diagrams of YIG/Py and single YIG structures. (c) Diagram of spin wave propagation in YIG/Py structure.

Based on the research of Zhang et al., Lei Zheng et al. proposed a design of a broadband XOR logic gate, as shown in Fig. 26(a) [40]. The structure is composed of input waveguide, output waveguide and interference Y-type waveguide. Due to the effect of dipole coupled, the wavelength of the spin wave on the side close to Co increases locally. Therefore, π phase shift can be achieved by setting appropriate structural parameters. Based on this feature, this structure can realize the function of a spin-wave XOR logic gate. The local inhomogeneity of the effective field on one side of the YIG waveguide is the main reason for the phase shift. The stray field of the Co element causes this local inhomogeneity. Normally, a fixed structure corresponds to a specific operating frequency. The magnitude of the dipole field is related to the magnetization angle between the materials. The magnitude of dipole field is directly related to the operating frequency of XOR gate. Therefore, Lei Zheng et al. used the SOT effect to adjust the magnetization direction of Co, thereby adjusting the dipole coupled between Co and YIG. Fig. 26 (c) and (d) shows the results of the SOT effect modulation. This method uses the edge-mode spin waves in the same waveguide to achieve spin-wave interference, and finally realizes the function of XOR logic gate. This provides a new way for the design of spin wave devices. But the current required to expand its operating frequency is relatively large because of the existence of external bias field. The existence of an external bias field will not only increase the size of the device, but also make the integration of the device very difficult. In addition, the existence of external bias field will increase the power consumption of device modulation. For spin wave devices, DE waves or surface waves are the typical choices. However, in order to achieve DE or surface spin waves (M⊥k), the waveguide needs to be magnetized along its short (hard) axis.

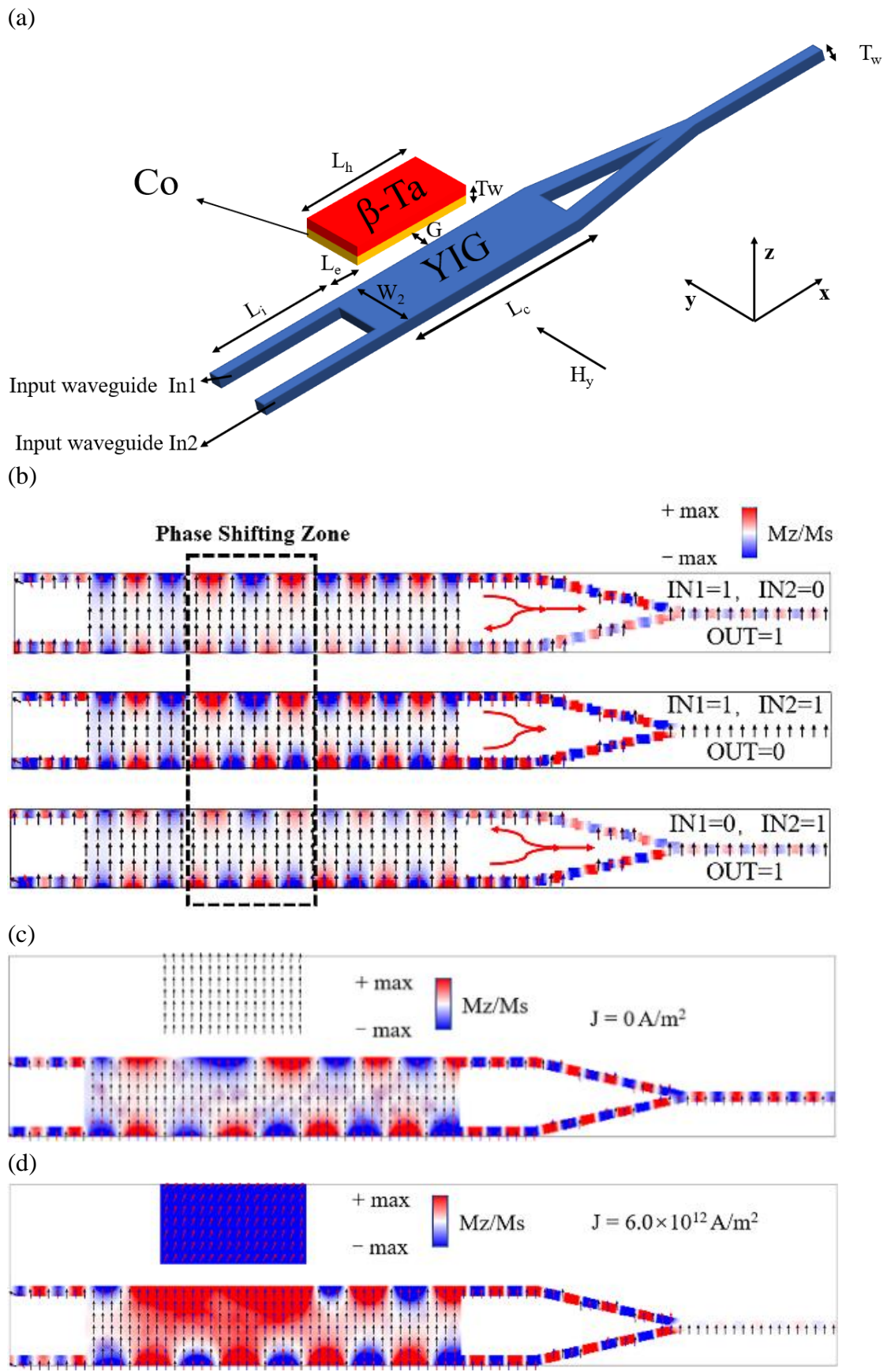

Fig. 26 (a) Schematic of SOT-based broadband spin wave XOR logic gate. (b) The simulated spatial maps of the logic gate under all logic inputs. (c) Diagram of spin wave propagation with current density of J=0 A/m² at 4.41GHz excitation frequency. (d) Diagram of spin wave propagation with current density of J=6.0×10¹² A/m² under 4.41GHz excitation frequency.

(a)

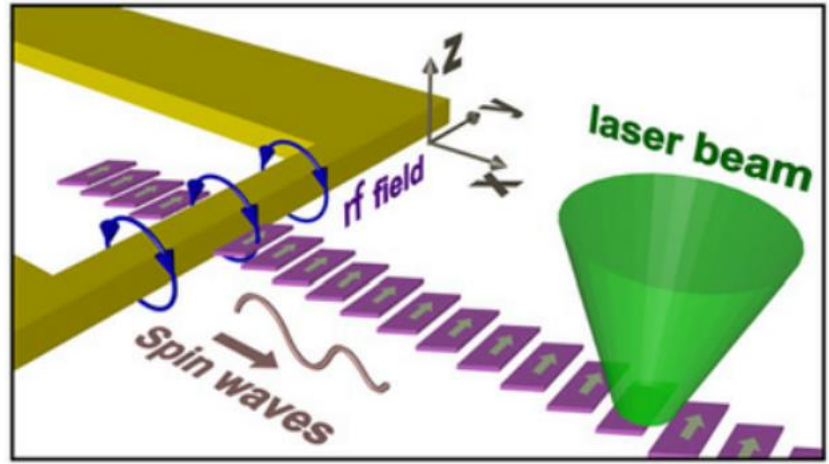

(b)

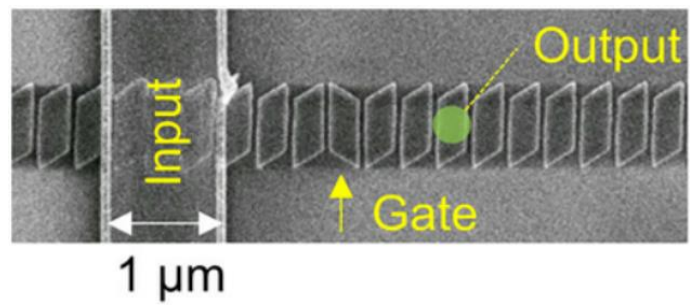

(c)

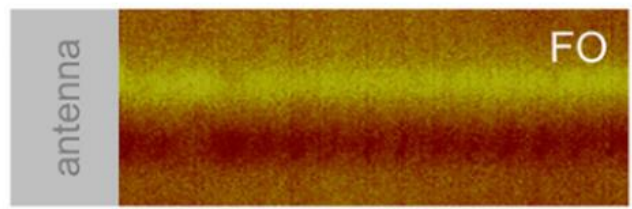

(d)

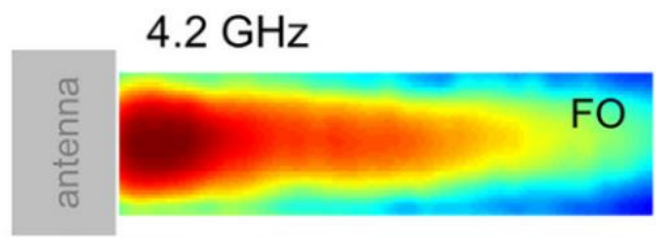

Fig. 27 (a) Schematic of RNM based waveguide. (b) SEM image of the RNM waveguide. (c) MFM image showing all the RNMs point in the same direction (FO state). (d) 2D spatial profile of the spin wave intensity for the FO state.

In order to solve this problem, Arabinda Haldar et al. proposed the waveguide structure composed of nanomagnets[41], as shown in Fig. 27 (a) and (b). This special type waveguide is designed based on the dipole-coupled but physically separated rhomboid-shaped nanomagnets (RNMs). Utilizing the dipole coupled between nanomagnets, the nanomagnets form the waveguide. The shape anisotropy of the nanomagnet is used to make its magnetization direction along the short axis of the waveguide. This structure solves the problem of the bias field cleverly. Arabinda Haldar et al. performed spin wave excitation and detection in this structure. Before spin wave excitation, in order to make all nanomagnets have the same magnetization direction (ferromagnetically ordered (FO) state), Arabinda Haldar et al. initialized all nanomagnets along the long axis, and then removed the magnetic field, as shown in Fig. 27 (c). By using $\mu$-BLS technology, an obvious propagating spin wave mode is observed at 4.2 GHz, as shown in Fig. 27 (d). Although this method avoids the problem of constantly applying a bias magnetic field during the use of the device, it still needs a bias magnetic field during initial magnetization. This type of waveguide provides a new idea for the integration of spin wave devices. However, the existence of the initializing magnetic field and the propagation efficiency of the spin wave between nanomagnets are still problems that need to be solved urgently by such methods.

This section focuses on some methods of modulating spin waves based on dipole coupled. These methods indirectly change the propagation characteristics of the spin wave through dipole field. This kind of method can be used not only to design fixed function devices, but also to realize reconfigurable logic devices by changing the magnetic state of resonators or placing materials with high saturation magnetization. In addition, the dipole coupled effect also plays an important role in self-biased waveguides. However, the problem with this type of method lies in the power consumption of the device and the propagation efficiency of the spin wave.

## 4. Summary

Spin waves have the potential to play a major role in the future of the microelectronics industry because of its low energy dissipation and high frequency. The research on the propagation characteristics of spin waves in uniform waveguide has experienced tremendous progress in the past few decades. Based on these research advances, researchers have come up with many efficient spin wave modulation methods, such as magnetic structure modulation and spin-orbit torque modulation. Obviously, the process required from a single basic spin wave device to a circuit and then to a system is very large. Fortunately, using SOT or STT to modulate the spin wave makes it possible for the compatibility between the spin wave devices and CMOS devices. In addition, the research on CMOS-based majority gates has been conducted for decades, but the efficiency of CMOS-based majority gates is still very low. For the spin wave devices, the logic structure to realize majority gates is simpler. Therefore, the research on the compatibility and complementarity between spin wave devices and CMOS devices is the main development trend.

The research on the propagation characteristics and modulation methods of spin-waves has brought about development of spin wave devices with initial achievement. Spin wave devices have obvious advantages over CMOS devices in terms of energy dissipation, and logic structure. However, the integration of spin wave devices and the realization of compatibility with CMOS circuits are still facing great challenges. The phase shifter based on spin wave is the main part of spin wave logic devices. As mentioned in Sec. III, there are many ways to realize spin wave phase shifter. Although methods such as STT or SOT have greatly improved the efficiency of spin wave phase shifting, improving the phase-shifting efficiency and reducing the dependence of external magnetic field are the urgent problems to be solved in the future for it. In addition, the spin wave coupler plays an important role in the connection of spin wave logic devices. As mentioned in Sec. II, spin waves propagate in different structure waveguides, there will be different modes and some special transmission phenomena. Although the spin wave coupler has made great progress in recent years, it still faces great challenges in cascade matching, signal transmission loss and applicable frequency band. For the application of spin wave devices, the devices based on CMOS are relatively mature and have a high market share. Therefore, spin wave devices should be developed towards compatibility with CMOS devices. With the development of spin wave theory and related process technology, we believe that the integration of spin wave circuits will soon be realized.

# Acknowledgements

This work was supported by the National Natural Science Foundation of China under grant Nos. 61734002, 62171079 and 51827802, the National Key Research and Development Plan (No. 2016YFA0300801), and the Sichuan Science and Technology Support Project (No. 2017JY0002).

# Data Availability

The data that support the findings of this study are available from the corresponding author upon reasonable request.